# A Quasi-Conforming Embedded Reproducing Kernel Particle Method for Heterogeneous Materials

Ryan T. Schlinkman[1], Jonghyuk Baek[1], Frank N. Beckwith[2], Stacy M. Nelson[2], and J. S. Chen[1*]

[1]University of California, San Diego

[2]Sandia National Laboratories, Albuquerque, NM

## Abstract

We present a quasi-conforming embedded reproducing kernel particle method (QCE-RKPM) for modeling heterogeneous materials that makes use of techniques not available to mesh-based methods such as the finite element method (FEM) and avoids many of the drawbacks in current embedded and immersed formulations which are based on meshed methods. The different material domains are discretized independently thus avoiding time-consuming, conformal meshing. In this approach, the superposition of foreground (inclusion) and background (matrix) domain integration smoothing cells are corrected by a quasi-conforming quadtree subdivision on the background integration smoothing cells. Due to the non-conforming nature of the background integration smoothing cells near the material interfaces, a variationally consistent (VC) correction for domain integration is introduced to restore integration constraints and thus optimal convergence rates at a minor computational cost. Additional interface integration smoothing cells with area (volume) correction, while non-conforming, can be easily introduced to further enhance the accuracy and stability of the Galerkin solution using VC integration on non-conforming cells. To properly approximate the weak discontinuity across the material interface by a penalty-free Nitsche's method with enhanced coercivity, the interface nodes on the surface of the foreground discretization are also shared with the background discretization. As such, there are no tunable parameters, such as those involved in the penalty type method, to enforce interface compatibility in this approach. The advantage of this meshfree formulation is that it avoids many of the instabilities in mesh-based immersed and embedded methods. The effectiveness of QCE-RKPM is illustrated with several examples.

## 1. Introduction

A vast array of engineering problems involving heterogeneous materials have been modelled by the finite element method (FEM) with a body-fitted mesh in which the discretizations of two different materials conform to their interface allowing for gradient jumps across the interface. Constructing such meshes, however, can be quite time-consuming and complicated especially for complex, three-dimensional geometry. One way to reduce meshing time for heterogeneous materials is to use embedded or immersed methods, both of which allow one to create independent, typically uniform discretizations of the different materials and which can sort out the communication between the meshes during the analysis. However, this reduction in manual meshing time comes at the cost of the algorithmic complexity in enforcing the constraint between the two materials while maintaining stability and accuracy.

“Immersed methods” refer to those methods in which the weak form equations assume overlapping meshes and, therefore, a fictitious background domain is placed underneath the foreground domain.

These may be classified by the constraint type with 1) interface-constrained immersed methods (ICIMs) enforcing compatibility only on the interface and 2) volume-constrained immersed methods (VCIMs) constraining the background deformation in the entire overlapping region to the foreground motion. The first ICIM devised was the immersed boundary method (IBM) proposed by Peskin [1, 2] (cf. [3, 4, 5] for FEM and hybrid versions) which distributes fluid-structure interaction (FSI) forces and interpolates velocities using a smoothed approximation of the Dirac delta function. However, IBM severely restricts the mesh size ratio between foreground and background to prevent fluid from leaking through the solid boundary [4]. Also, the smooth Dirac delta function smears the solution over several grid cells rather than a sharp discontinuity in the normal gradient at the interface thus limiting IBM to first-order accuracy unless more complicated procedures are added [6, 7]. A more recent ICIM formulation introduced by Bazilevs et al. [8, 9] uses a uniform background mesh discretized with IGA splines [10] to approximate the momentum equations for both the background and foreground meshes. One issue with this approach is the undesirable smooth background approximation for the discontinuous kinematics in fracture and fragmentation which results in damage zones whose size scales with the background mesh [11].

VCIMs began with Glowinsky et al.'s distributed Lagrange multiplier method [12] which used Lagrange multipliers (LM) requiring the approximation spaces to meet the LBB condition [13, 14]. Developments of VCIMs in recent years include the immersed finite element method (IFEM) [15, 16] and the modified IFEM (mIFEM) [17] which use the reproducing kernel (RK) approximation [18, 19] to interpolate velocities between solid and fluid, and FSI forces are treated as volumetric body forces. However, like all VCIMs, the smooth approximation in the fluid smears the effect of the interface over several elements leading to low convergence rates [20]. The variational multiscale immersed method (VMIM) for heterogeneous materials [21] uses the reproducing kernel particle method (RKPM) with volumetric Lagrange multipliers and a variational multiscale decomposition [22, 23] of the background approximation leading to a residual-based stabilization which also suppresses the leaking instability caused by large solid to fluid mesh size ratios [24]. But again, it too suffers from large interpolation errors near the interface resulting in low convergence rates [21]. While VCIMs are effective without a tedious contour integral along the interface, achieving optimal convergence rates requires a direct treatment of the gradient discontinuity at the interface [25]. Furthermore, VCIMs require some sort of stabilization [21, 26].

“Embedded methods” refer to those methods in which there are no overlapping meshes, and the conformity of matrix and inclusion domains (which allows for discontinuous normal gradients) is solved using techniques such as cutting elements, aggregating elements, using surrogate domains, etc. Cut element approaches either add extra DOFs to elements intersecting the interface (e.g., CutFEM [27, 28]) or divide the original elements into smaller “integration elements” (e.g., discontinuity-enriched FEM (DE-FEM) [29, 30, 31]). Nonetheless, cutting or dividing elements requires potentially complex computational geometry algorithms and, more importantly, creates what is called the small cut element problem in which an element can be cut in an arbitrary fashion leaving a very small or slender element on one side [32]. This results in a very small critical time step and an ill-conditioned stiffness matrix. Several methods have been proposed to mitigate the effect of the cut cell problem such as the ghost penalty [32, 33, 34], bubble functions [35, 36], carefully weighting the gradients in the Nitsche integrals [37, 38], or discarding the problematic DOFs altogether [31]. The Finite Cell Method (FCM) [39, 40, 41, 42] utilizes an adaptive integration scheme which sub-divides elements into sub-cells (not sub-elements) to place quadrature points thus allowing one to better approximate the boundary/interface and avoiding the problem of small elements. However, one must divide elements many times to accurately approximate the domain being integrated which yields a computationally intensive domain integration [41].

The shifted boundary/interface method (SBM/SIM) [43, 44, 45] integrates over a surrogate domain (e.g., all elements wholly inside the interface) and shifts the interface conditions from the true interface to the surrogate interface using a Taylor series expansion involving the intersected elements' gradients; this allows the method to maintain optimal rates of convergence without cutting elements (cf. [46, 47] for a similar idea). However, the transfer algorithm involves a potentially complex closest point projection procedure, and the transfer of Neumann boundary conditions requires either a higher-order approximation [48] or a stabilized equal-order mixed method [49] to prevent reduction in convergence rates. In the aggregated finite element method (AgFEM) [50, 51, 52], elements intersected by the interface are merged with elements inside the interface, and the problematic DOFs are constrained as a linear combination of non-problematic DOFs (cf. similar approaches in [53, 54, 55, 56]). This avoids small cut elements and achieves optimal convergence rates but can possibly introduce error in the aggregation and constraint process.

In this paper, we propose a formulation that fits into the category of embedded methods but is unique in that it makes use of techniques available exclusively to meshfree methods and avoids many of the pitfalls of the formulations outlined above due to their reliance on mesh-based methods. First, the background (matrix) and foreground (inclusions) are discretized independently with superposition. Then what follows is a quasi-conforming quadtree subdivision on the background integration smoothing cells where needed (which mesh-based methods are incapable of doing). Due to the loss of integration smoothing cell conformity, a variationally consistent (VC) correction [57] of domain integration is employed at a minimal computational cost to restore the integration constraint for Galerkin exactness, enhance accuracy, and achieve optimal convergence rates. The interface nodes are shared by both domains which provides enough coercivity to make Nitsche's penalty parameter unnecessary. Thus, unlike most immersed and embedded methods, this formulation does not have any tunable parameters present to enforce interface compatibility. Our meshfree method of choice is the reproducing kernel particle method (RKPM), and so, we call this formulation a quasi-conforming embedded RKPM (QCE-RKPM).

The remaining parts of this paper are outlined as follows. In section 2, we detail the equations for heterogeneous materials. Section 3 reviews the reproducing kernel approximation and discusses domain integration in meshfree methods. Section 4 details the discretization process of the background domain from an immersed state to a quasi-conforming embedded domain. Section 5 then presents numerical examples to demonstrate the effectiveness of our method, and section 6 summarizes our findings.

## 2. Governing Equations

### 2.1. Strong Form

Consider a heterogeneous solid domain $\Omega$ comprised of a matrix domain $\Omega^-$ and an inclusion domain $\Omega^+$, where $\overline{\Omega} = \overline{\Omega}^+ \cup \overline{\Omega}^-$. The interface between the two materials is denoted by $\Gamma$, where $\Gamma = \overline{\Omega}^+ \cap \overline{\Omega}^-$. The matrix material possesses an outer boundary $\partial\Omega$ which is comprised of Neumann $\partial\Omega_N$ and Dirichlet $\partial\Omega_D$ components such that $\partial\Omega = \partial\Omega_N \cup \partial\Omega_D$ and $\partial\Omega_N \cap \partial\Omega_D = \emptyset$, see Figure 1(a) for an illustration.

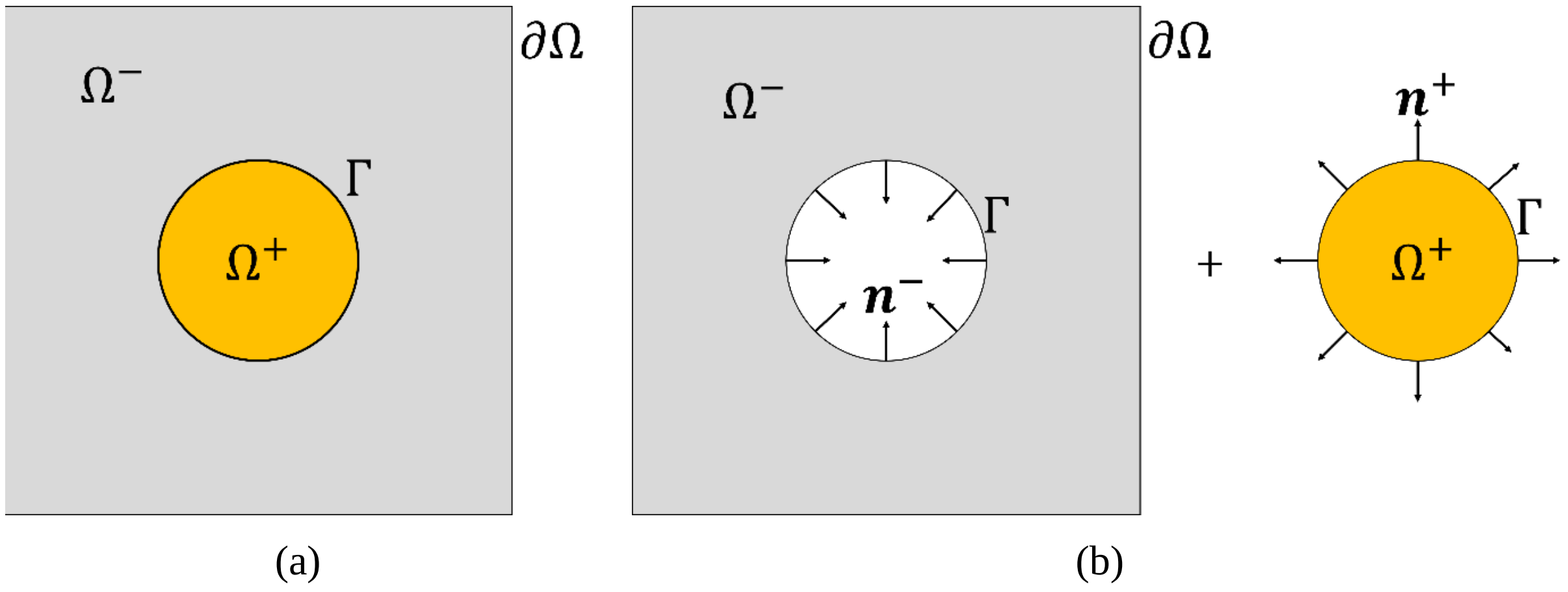

Figure 1: (a) Heterogeneous material domain consisting of matrix $\Omega^-$ and inclusion $\Omega^+$ materials with outer boundary $\partial\Omega$ and interface $\Gamma$ between the two materials and (b) separate matrix and inclusion domains with their outward normal vectors on the interface

Our governing equation is the equation of equilibrium for a linear elastic material:

$$\nabla \cdot \boldsymbol{\sigma}(\boldsymbol{x}) + \boldsymbol{b}(\boldsymbol{x}) = \boldsymbol{0}, \qquad \boldsymbol{x} \in \Omega, \tag{1}$$

$$\boldsymbol{\sigma}(\boldsymbol{x}) \cdot \boldsymbol{n}(\boldsymbol{x}) = \boldsymbol{t}(\boldsymbol{x}), \qquad \boldsymbol{x} \in \partial\Omega_N \tag{2}$$

$$\boldsymbol{u}(\boldsymbol{x}) = \boldsymbol{g}(\boldsymbol{x}), \qquad \boldsymbol{x} \in \partial\Omega_D, \tag{3}$$

$$\boldsymbol{\sigma}(\boldsymbol{x}) = \boldsymbol{C}(\boldsymbol{x}) : \boldsymbol{\varepsilon}\big(\boldsymbol{u}(\boldsymbol{x})\big), \qquad \boldsymbol{x} \in \Omega, \tag{4}$$

where:

$$\boldsymbol{b}(\boldsymbol{x}) = \begin{cases} \boldsymbol{b}^-, & \boldsymbol{x} \in \Omega^- \\ \boldsymbol{b}^+, & \boldsymbol{x} \in \Omega^+ \end{cases}, \tag{5}$$

$$\boldsymbol{u}(\boldsymbol{x}) = \begin{cases} \boldsymbol{u}^-, & \boldsymbol{x} \in \Omega^- \\ \boldsymbol{u}^+, & \boldsymbol{x} \in \Omega^+ \end{cases}, \tag{6}$$

$$\boldsymbol{C}(\boldsymbol{x}) = \begin{cases} \boldsymbol{C}^-, & \boldsymbol{x} \in \Omega^- \\ \boldsymbol{C}^+, & \boldsymbol{x} \in \Omega^+ \end{cases}, \tag{7}$$

hence:

$$\boldsymbol{\sigma}(\boldsymbol{x}) = \begin{cases} \boldsymbol{\sigma}^-(\boldsymbol{u}^-) = \boldsymbol{C}^-:\boldsymbol{\varepsilon}(\boldsymbol{u}^-), & \boldsymbol{x} \in \Omega^- \\ \boldsymbol{\sigma}^+(\boldsymbol{u}^+) = \boldsymbol{C}^+:\boldsymbol{\varepsilon}(\boldsymbol{u}^+), & \boldsymbol{x} \in \Omega^+ \end{cases}, \tag{8}$$

where $\boldsymbol{\sigma}(\boldsymbol{x})$ is the Cauchy stress tensor, $\boldsymbol{b}(\boldsymbol{x})$ is the body force vector, $\boldsymbol{n}(\boldsymbol{x})$ is the outward normal vector on $\partial\Omega_N$, $\boldsymbol{t}(\boldsymbol{x})$ is the prescribed traction vector on $\partial\Omega_N$, $\boldsymbol{u}(\boldsymbol{x})$ is the displacement vector, $\boldsymbol{g}(\boldsymbol{x})$ is the prescribed displacement vector on $\partial\Omega_D$, $\boldsymbol{C}(\boldsymbol{x})$ is the elastic constitutive tensor, and $\boldsymbol{\varepsilon}(\boldsymbol{u}(\boldsymbol{x})) = \frac{1}{2}(\nabla\otimes\boldsymbol{u}(\boldsymbol{x}) + \boldsymbol{u}(\boldsymbol{x})\otimes\nabla)$ is the small strain tensor. Additionally, continuity and equilibrium conditions on the interface $\Gamma$ are:

$$\boldsymbol{u}^+ - \boldsymbol{u}^- = \boldsymbol{0}, \qquad \forall\boldsymbol{x} \in \Gamma, \tag{9}$$

$$\boldsymbol{\sigma}^+(\boldsymbol{u}^+)\cdot\boldsymbol{n}^+ + \boldsymbol{\sigma}^-(\boldsymbol{u}^-)\cdot\boldsymbol{n}^- = \boldsymbol{0}, \qquad \forall\boldsymbol{x} \in \Gamma, \tag{10}$$

where $\boldsymbol{n}^+$ and $\boldsymbol{n}^-$ are the outward normal vectors on $\Gamma$ from the inclusion and matrix sides, respectively, as illustrated in Figure 1(b).

### 2.2. Weak Form

To begin deriving the weak form of Eqs. (1) through (4), we multiply Eq. (1) by a test function $\delta\boldsymbol{u}(\boldsymbol{x})$ and integrate over the domain $\Omega$:

$$\left(\delta\boldsymbol{u}(\boldsymbol{x}), \nabla\cdot\boldsymbol{\sigma}(\boldsymbol{x})\right)_\Omega + \left(\delta\boldsymbol{u}(\boldsymbol{x}), \boldsymbol{b}(\boldsymbol{x})\right)_\Omega = \boldsymbol{0}, \qquad \boldsymbol{x} \in \Omega, \tag{11}$$

where $(\cdot,\cdot)_\Omega$ is the inner product in $\Omega$, and

$$\delta\boldsymbol{u}(\boldsymbol{x}) = \begin{cases} \delta\boldsymbol{u}^-, & \boldsymbol{x} \in \Omega^- \\ \delta\boldsymbol{u}^+, & \boldsymbol{x} \in \Omega^+ \end{cases}. \tag{12}$$

Given the definitions in Eqs. (5) through (8) and (12), Eq. (11) may be split by domain:

$$\left(\delta\boldsymbol{u}^-, \nabla\cdot\boldsymbol{\sigma}^-(\boldsymbol{u}^-)\right)_{\Omega^-} + (\delta\boldsymbol{u}^-, \boldsymbol{b}^-)_{\Omega^-} = \mathbf{0}, \tag{13}$$

$$\left(\delta\boldsymbol{u}^+, \nabla\cdot\boldsymbol{\sigma}^+(\boldsymbol{u}^+)\right)_{\Omega^+} + (\delta\boldsymbol{u}^+, \boldsymbol{b}^+)_{\Omega^+} = \mathbf{0}. \tag{14}$$

Performing integration by parts on the terms containing the stress divergences and using Gauss's theorem, we have:

$$\left(\boldsymbol{\varepsilon}(\delta\boldsymbol{u}^-), \boldsymbol{\sigma}^-(\boldsymbol{u}^-)\right)_{\Omega^-} - \left(\delta\boldsymbol{u}^-, \boldsymbol{n}^-\cdot\boldsymbol{\sigma}^-(\boldsymbol{u}^-)\right)_{\Gamma} = (\delta\boldsymbol{u}^-, \boldsymbol{b}^-)_{\Omega^-} + (\delta\boldsymbol{u}^-, \boldsymbol{t})_{\partial\Omega_N}, \tag{15}$$

$$\left(\boldsymbol{\varepsilon}(\delta\boldsymbol{u}^+), \boldsymbol{\sigma}^+(\boldsymbol{u}^+)\right)_{\Omega^+} - \left(\delta\boldsymbol{u}^+, \boldsymbol{n}^+\cdot\boldsymbol{\sigma}^+(\boldsymbol{u}^+)\right)_{\Gamma} = (\delta\boldsymbol{u}^+, \boldsymbol{b}^+)_{\Omega^+}. \tag{16}$$

Note that by substituting $\boldsymbol{n}^- = -\boldsymbol{n}^+$ into interface equilibrium (10), we have:

$$\boldsymbol{\sigma}^+(\boldsymbol{u}^+)\cdot\boldsymbol{n}^+ = \boldsymbol{\sigma}^-(\boldsymbol{u}^-)\cdot\boldsymbol{n}^+. \tag{17}$$

This implies that $\boldsymbol{\sigma}^+(\boldsymbol{u}^+)$, $\boldsymbol{\sigma}^-(\boldsymbol{u}^-)$, or a linear combination of the two may be used interchangeably in the interface traction terms. Define a linear combination of stresses on the interface as:

$$\overline{\boldsymbol{\sigma}}(\boldsymbol{u}) \equiv \alpha\boldsymbol{\sigma}^+(\boldsymbol{u}^+) + (1-\alpha)\boldsymbol{\sigma}^-(\boldsymbol{u}^-) \tag{18}$$

where $\alpha \in [0,1]$. We can rearrange (18) as $\overline{\boldsymbol{\sigma}}(\boldsymbol{u}) = \boldsymbol{\sigma}^-(\boldsymbol{u}^-) + \alpha\left(\boldsymbol{\sigma}^+(\boldsymbol{u}^+) - \boldsymbol{\sigma}^-(\boldsymbol{u}^-)\right)$, and using (17), we have:

$$\overline{\boldsymbol{\sigma}}(\boldsymbol{u})\cdot\boldsymbol{n}^+ = \boldsymbol{\sigma}^+(\boldsymbol{u}^+)\cdot\boldsymbol{n}^+ = -\boldsymbol{\sigma}^-(\boldsymbol{u}^-)\cdot\boldsymbol{n}^-. \tag{19}$$

Substituting (19) into (15) and (16), we have:

$$\left(\boldsymbol{\varepsilon}(\delta\boldsymbol{u}^-),\boldsymbol{\sigma}^-(\boldsymbol{u}^-)\right)_{\Omega^-}+\left(\delta\boldsymbol{u}^-,\boldsymbol{n}^+\cdot\overline{\boldsymbol{\sigma}}(\boldsymbol{u})\right)_\Gamma=(\delta\boldsymbol{u}^-,\boldsymbol{b}^-)_{\Omega^-}+(\delta\boldsymbol{u}^-,\boldsymbol{t})_{\partial\Omega_N}, \tag{20}$$

$$\left(\boldsymbol{\varepsilon}(\delta\boldsymbol{u}^+),\boldsymbol{\sigma}^+(\boldsymbol{u}^+)\right)_{\Omega^+}-\left(\delta\boldsymbol{u}^+,\boldsymbol{n}^+\cdot\overline{\boldsymbol{\sigma}}(\boldsymbol{u})\right)_\Gamma=(\delta\boldsymbol{u}^+,\boldsymbol{b}^+)_{\Omega^+}, \tag{21}$$

or:

$$\begin{aligned}&\left(\boldsymbol{\varepsilon}(\delta\boldsymbol{u}^-),\boldsymbol{\sigma}^-(\boldsymbol{u}^-)\right)_{\Omega^-}+\alpha\left(\delta\boldsymbol{u}^-,\boldsymbol{n}^+\cdot\boldsymbol{\sigma}^+(\boldsymbol{u}^+)\right)_\Gamma+(1-\alpha)\left(\delta\boldsymbol{u}^-,\boldsymbol{n}^+\cdot\boldsymbol{\sigma}^-(\boldsymbol{u}^-)\right)_\Gamma\\&\quad=(\delta\boldsymbol{u}^-,\boldsymbol{b}^-)_{\Omega^-}+(\delta\boldsymbol{u}^-,\boldsymbol{t})_{\partial\Omega_N},\end{aligned} \tag{22}$$

$$\begin{aligned}&\left(\boldsymbol{\varepsilon}(\delta\boldsymbol{u}^+),\boldsymbol{\sigma}^+(\boldsymbol{u}^+)\right)_{\Omega^+}-\alpha\left(\delta\boldsymbol{u}^+,\boldsymbol{n}^+\cdot\boldsymbol{\sigma}^+(\boldsymbol{u}^+)\right)_\Gamma-(1-\alpha)\left(\delta\boldsymbol{u}^+,\boldsymbol{n}^+\cdot\boldsymbol{\sigma}^-(\boldsymbol{u}^-)\right)_\Gamma\\&\quad=(\delta\boldsymbol{u}^+,\boldsymbol{b}^+)_{\Omega^+},\end{aligned} \tag{23}$$

Since meshfree shape functions do not possess the Kronecker delta property, we use Nitsche's method to enforce Dirichlet boundary conditions. Our weak form reads: Find $(\boldsymbol{u}^+,\boldsymbol{u}^-)\in H^1(\Omega^+)\times H^1(\Omega^-)$ such that $\forall(\delta\boldsymbol{u}^+,\delta\boldsymbol{u}^-)\in H^1(\Omega^+)\times H^1(\Omega^-)$

$$\begin{aligned}&\left(\boldsymbol{\varepsilon}(\delta\boldsymbol{u}^-),\boldsymbol{\sigma}^-(\boldsymbol{u}^-)\right)_{\Omega^-}+\alpha\left(\delta\boldsymbol{u}^-,\boldsymbol{n}^+\cdot\boldsymbol{\sigma}^+(\boldsymbol{u}^+)\right)_\Gamma+(1-\alpha)\left(\delta\boldsymbol{u}^-,\boldsymbol{n}^+\cdot\boldsymbol{\sigma}^-(\boldsymbol{u}^-)\right)_\Gamma\\&\quad-\left(\delta\boldsymbol{u}^-,\boldsymbol{n}^-\cdot\boldsymbol{\sigma}^-(\boldsymbol{u}^-)\right)_{\partial\Omega_D}-(\boldsymbol{\sigma}^-(\delta\boldsymbol{u}^-)\cdot\boldsymbol{n}^-,\boldsymbol{u}^-)_{\partial\Omega_D}+\beta(\delta\boldsymbol{u}^-,\boldsymbol{u}^-)_{\partial\Omega_D}\\&\quad=(\delta\boldsymbol{u}^-,\boldsymbol{b}^-)_{\Omega^-}+(\delta\boldsymbol{u}^-,\boldsymbol{t})_{\partial\Omega_N}-(\boldsymbol{\sigma}^-(\delta\boldsymbol{u}^-)\cdot\boldsymbol{n}^-,\boldsymbol{g})_{\partial\Omega_D}+\beta(\delta\boldsymbol{u}^-,\boldsymbol{g})_{\partial\Omega_D},\end{aligned} \tag{24}$$

$$\begin{aligned}&\left(\boldsymbol{\varepsilon}(\delta\boldsymbol{u}^+),\boldsymbol{\sigma}^+(\boldsymbol{u}^+)\right)_{\Omega^+}-\alpha\left(\delta\boldsymbol{u}^+,\boldsymbol{n}^+\cdot\boldsymbol{\sigma}^+(\boldsymbol{u}^+)\right)_\Gamma-(1-\alpha)\left(\delta\boldsymbol{u}^+,\boldsymbol{n}^+\cdot\boldsymbol{\sigma}^-(\boldsymbol{u}^-)\right)_\Gamma\\&\quad=(\delta\boldsymbol{u}^+,\boldsymbol{b}^+)_{\Omega^+},\end{aligned} \tag{25}$$

where $\beta$ is the penalty parameter.

**Remark 2.1**

*Alternatively, one could derive a symmetric weak form from the following potential:*

$$\begin{aligned}\Pi = &\frac{1}{2}\big(\boldsymbol{\varepsilon}(\boldsymbol{u}^-), \boldsymbol{C}^-:\boldsymbol{\varepsilon}(\boldsymbol{u}^-)\big)_{\Omega^-} + \frac{1}{2}\big(\boldsymbol{\varepsilon}(\boldsymbol{u}^+), \boldsymbol{C}^+:\boldsymbol{\varepsilon}(\boldsymbol{u}^+)\big)_{\Omega^+} - (\boldsymbol{u}^-, \boldsymbol{b}^-)_{\Omega^-} \\ &- (\boldsymbol{u}^-, \boldsymbol{t})_{\partial\Omega_N} - (\boldsymbol{u}^+, \boldsymbol{b}^+)_{\Omega^+} - (\overline{\boldsymbol{\sigma}}(\boldsymbol{u}) \cdot \boldsymbol{n}^+, \boldsymbol{u}^+ - \boldsymbol{u}^-)_\Gamma \\ &- (\boldsymbol{\sigma}^-(\boldsymbol{u}^-) \cdot \boldsymbol{n}^-, \boldsymbol{u}^- - \boldsymbol{g})_{\partial\Omega_D} + \frac{\beta}{2}(\boldsymbol{u}^- - \boldsymbol{g}, \boldsymbol{u}^- - \boldsymbol{g})_{\partial\Omega_D}.\end{aligned} \tag{26}$$

*Taking the variation of the potential, setting it equal to zero, splitting the equation in two by test function, and rearranging, we have:*

$$\begin{aligned}&\big(\boldsymbol{\varepsilon}(\delta\boldsymbol{u}^-), \boldsymbol{\sigma}^-(\boldsymbol{u}^-)\big)_{\Omega^-} + \alpha\big(\delta\boldsymbol{u}^-, \boldsymbol{n}^+ \cdot \boldsymbol{\sigma}^+(\boldsymbol{u}^+)\big)_\Gamma + (1-\alpha)\big(\delta\boldsymbol{u}^-, \boldsymbol{n}^+ \cdot \boldsymbol{\sigma}^-(\boldsymbol{u}^-)\big)_\Gamma \\ &\quad + (1-\alpha)(\boldsymbol{\sigma}^-(\delta\boldsymbol{u}^-) \cdot \boldsymbol{n}^+, \boldsymbol{u}^- - \boldsymbol{u}^+)_\Gamma - \big(\delta\boldsymbol{u}^-, \boldsymbol{n}^- \cdot \boldsymbol{\sigma}^-(\boldsymbol{u}^-)\big)_{\partial\Omega_D} \\ &\quad - (\boldsymbol{\sigma}^-(\delta\boldsymbol{u}^-) \cdot \boldsymbol{n}^-, \boldsymbol{u}^-)_{\partial\Omega_D} + \beta(\delta\boldsymbol{u}^-, \boldsymbol{u}^-)_{\partial\Omega_D} \\ &\quad = (\delta\boldsymbol{u}^-, \boldsymbol{b}^-)_{\Omega^-} + (\delta\boldsymbol{u}^-, \boldsymbol{t})_{\partial\Omega_N} - (\boldsymbol{\sigma}^-(\delta\boldsymbol{u}^-) \cdot \boldsymbol{n}^-, \boldsymbol{g})_{\partial\Omega_D} + \beta(\delta\boldsymbol{u}^-, \boldsymbol{g})_{\partial\Omega_D},\end{aligned} \tag{27}$$

$$\begin{aligned}&\big(\boldsymbol{\varepsilon}(\delta\boldsymbol{u}^+), \boldsymbol{\sigma}^+(\boldsymbol{u}^+)\big)_{\Omega^+} - \alpha\big(\delta\boldsymbol{u}^+, \boldsymbol{n}^+ \cdot \boldsymbol{\sigma}^+(\boldsymbol{u}^+)\big)_\Gamma - (1-\alpha)\big(\delta\boldsymbol{u}^+, \boldsymbol{n}^+ \cdot \boldsymbol{\sigma}^-(\boldsymbol{u}^-)\big)_\Gamma \\ &\quad + \alpha(\boldsymbol{\sigma}^+(\delta\boldsymbol{u}^+) \cdot \boldsymbol{n}^+, \boldsymbol{u}^- - \boldsymbol{u}^+)_\Gamma = (\delta\boldsymbol{u}^+, \boldsymbol{b}^+)_{\Omega^+}.\end{aligned} \tag{28}$$

*The only difference between this weak form and the one given in Eqs. (24) and (25) is the addition of the interface term* $-(\overline{\boldsymbol{\sigma}}(\delta\boldsymbol{u}) \cdot \boldsymbol{n}^+, \boldsymbol{u}^+ - \boldsymbol{u}^-)_\Gamma$. *This weak form is essentially Nitsche's method without penalty for enforcing interface compatibility. It has been shown in* [58] *that an interface penalty term is not needed if the interface nodes are shared by the foreground and background discretizations for enhanced coercivity.*

We have found that any value of $\alpha$ in Eqs. (24) and (25) using the proposed method to be discussed in Section 4 will provide an equally good approximation. This is in contrast with FEM cut-element techniques which must carefully balance the interface gradients to avoid matrix ill-conditioning and small critical time steps [37, 38]. For simplicity, we have chosen to use the inclusion interface stress exclusively for the interface terms (i.e., $\alpha = 1$). We thus arrive at the weak form used for our numerical examples:

$$
\begin{aligned}
&\big(\boldsymbol{\varepsilon}(\delta\boldsymbol{u}^-),\boldsymbol{\sigma}^-(\boldsymbol{u}^-)\big)_{\Omega^-}+\big(\delta\boldsymbol{u}^-,\boldsymbol{n}^+\cdot\boldsymbol{\sigma}^+(\boldsymbol{u}^+)\big)_{\Gamma}-\big(\delta\boldsymbol{u}^-,\boldsymbol{n}^-\cdot\boldsymbol{\sigma}^-(\boldsymbol{u}^-)\big)_{\partial\Omega_D}\\
&-(\boldsymbol{\sigma}^-(\delta\boldsymbol{u}^-)\cdot\boldsymbol{n}^-,\boldsymbol{u}^-)_{\partial\Omega_D}+\beta(\delta\boldsymbol{u}^-,\boldsymbol{u}^-)_{\partial\Omega_D}\\
&=(\delta\boldsymbol{u}^-,\boldsymbol{b}^-)_{\Omega^-}+(\delta\boldsymbol{u}^-,\boldsymbol{t})_{\partial\Omega_N}-(\boldsymbol{\sigma}^-(\delta\boldsymbol{u}^-)\cdot\boldsymbol{n}^-,\boldsymbol{g})_{\partial\Omega_D}+\beta(\delta\boldsymbol{u}^-,\boldsymbol{g})_{\partial\Omega_D},
\end{aligned}
\tag{29}
$$

$$
\big(\boldsymbol{\varepsilon}(\delta\boldsymbol{u}^+),\boldsymbol{\sigma}^+(\boldsymbol{u}^+)\big)_{\Omega^+}-\big(\delta\boldsymbol{u}^+,\boldsymbol{n}^+\cdot\boldsymbol{\sigma}^+(\boldsymbol{u}^+)\big)_{\Gamma}=(\delta\boldsymbol{u}^+,\boldsymbol{b}^+)_{\Omega^+}. \tag{30}
$$

# 3. Variationally Consistent Reproducing Kernel Particle Method

In this section, we detail the reproducing kernel approximation and discuss the stabilized nodal integration suitable for the proposed embedded method. The matrix form may be found in the appendix.

## 3.1. Reproducing Kernel Approximation

The reproducing kernel (RK) approximation [18, 19] is constructed based on a set of $NP$ scattered nodes that discretize the domain $\Omega$ as shown in Figure 2.

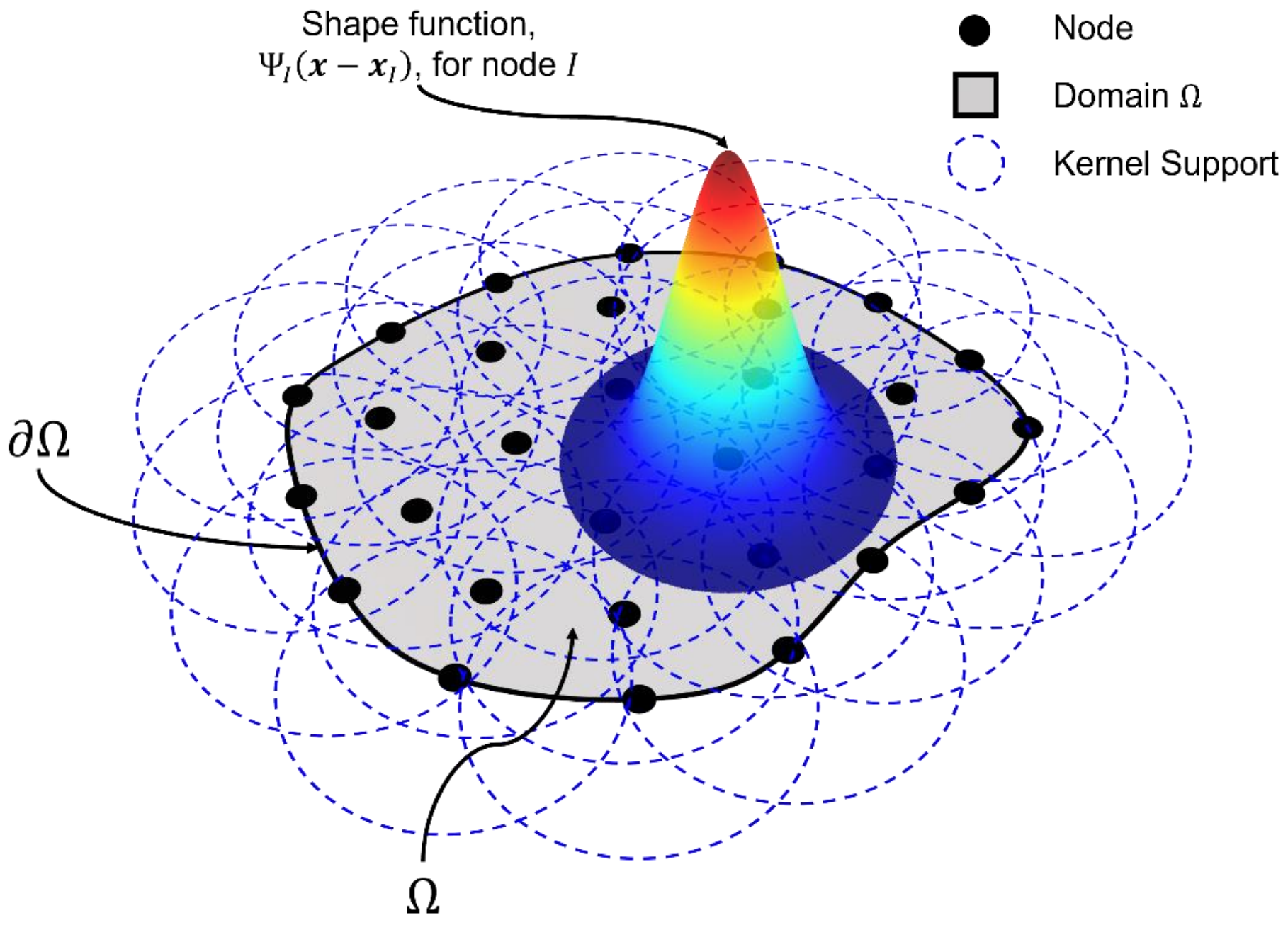

Figure 2: A domain $\Omega$ with boundary $\partial\Omega$ is discretized by a set of nodes. Each node possesses a circular kernel support where the shape function $\Psi_I$ is defined.

Let $u^h(\boldsymbol{x})$ be the approximation of $u(\boldsymbol{x})$; the RK approximation is given as follows [19]:

$$u^h(\boldsymbol{x}) = \sum_{I=1}^{NP} \Psi_I(\boldsymbol{x})\, u_I, \tag{31}$$

where $u_I$ is the generalized coefficient associated with node $I$, and $\Psi_I(\boldsymbol{x})$ is the RK approximation function associated with node $I$ expressed as:

$$\Psi_I(\boldsymbol{x}) = \mathbf{H}^T(\mathbf{0})\, \mathbf{M}^{-1}(\boldsymbol{x})\, \mathbf{H}(\boldsymbol{x} - \boldsymbol{x}_I)\, \phi_a(\boldsymbol{x} - \boldsymbol{x}_I), \tag{32}$$

where $\phi_a(\boldsymbol{x} - \boldsymbol{x}_I)$ is the kernel function that defines the locality and smoothness (i.e., the order of continuity) of the approximation. A common kernel function used in RK approximations is the cubic B-spline kernel which is $C^2$ continuous:

$$\phi_a(z) = \begin{cases} \frac{2}{3} - 4z_I^2 + 4z_I^3, & 0 \le z_I \le \frac{1}{2} \\ \frac{4}{3}(1 - z_I)^3, & \frac{1}{2} < z_I \le 1 \\ 0, & z_I > 1 \end{cases}, \qquad z_I = \frac{\|\boldsymbol{x} - \boldsymbol{x}_I\|}{a_I}, \tag{33}$$

and $\mathbf{M}(\boldsymbol{x})$ is the moment matrix:

$$\mathbf{M}(\boldsymbol{x}) = \sum_{I=1}^{NP} \mathbf{H}(\boldsymbol{x} - \boldsymbol{x}_I)\, \mathbf{H}^T(\boldsymbol{x} - \boldsymbol{x}_I)\, \phi_a(\boldsymbol{x} - \boldsymbol{x}_I), \tag{34}$$

where $\mathbf{H}(\boldsymbol{x} - \boldsymbol{x}_I)$ is the basis vector of monomials:

$$\mathbf{H}^T(\boldsymbol{x} - \boldsymbol{x}_I) = [1, x_1 - x_{1I}, x_2 - x_{2I}, x_3 - x_{3I}, (x_1 - x_{1I})^2, \dots, (x_3 - x_{3I})^n]. \tag{35}$$

Here $\mathbf{H}(\boldsymbol{x} - \boldsymbol{x}_I)$ of basis order $n$ enforces the $n^{\text{th}}$ order of completeness in the approximation:

$$\sum_{I=1}^{NP} \Psi_I(\boldsymbol{x})\, x_{1I}^i x_{2I}^j x_{3I}^k = x_1^i x_2^j x_3^k, \qquad 0 \le i + j + k \le n. \tag{36}$$

For the moment matrix to be invertible, the point being evaluated $\boldsymbol{x}$ must be covered by at least the number of non-collinear (in 2D) or non-coplanar (in 3D) kernel supports equal to the dimension of the basis vector [59].

**Remark 3.1**

1. *When using meshfree methods in the presence of concave features, which commonly exist in microstructures, it is necessary to use a line-of-sight algorithm to determine a node's influence on an evaluation point* [60, 61]. *The fact that a node's kernel support covers an evaluation point is not sufficient to determine influence; the node also needs to have a line* *of sight within the material body to the evaluation point. For example, consider Figure 3. Node 1's kernel support covers nodes 2, 3, and 4 along with other nodes in the domain. Node 1 has line-of-sight within the domain to node 2, and since the boundary of the domain connects nodes 1 and 3 with a straight line, node 1 also has line-of-sight to node 3. While node 4 is within node 1's kernel support, a straight line drawn from node 1 to node 4 would pass outside the domain, and thus node 1's kernel will have a value of zero when evaluated at node 4.*

2. *The line-of-sight algorithm creates a discontinuous kernel function, which leads to an undesired discontinuity in the approximation. Another approach to handle concave features while maintaining continuity in the approximation is to use a diffracted kernel support* [62, 61, 63] *which bends around the corner, e.g., at node 3 in Figure 3. In this work, we consider the line-of-sight approach for simplicity.*

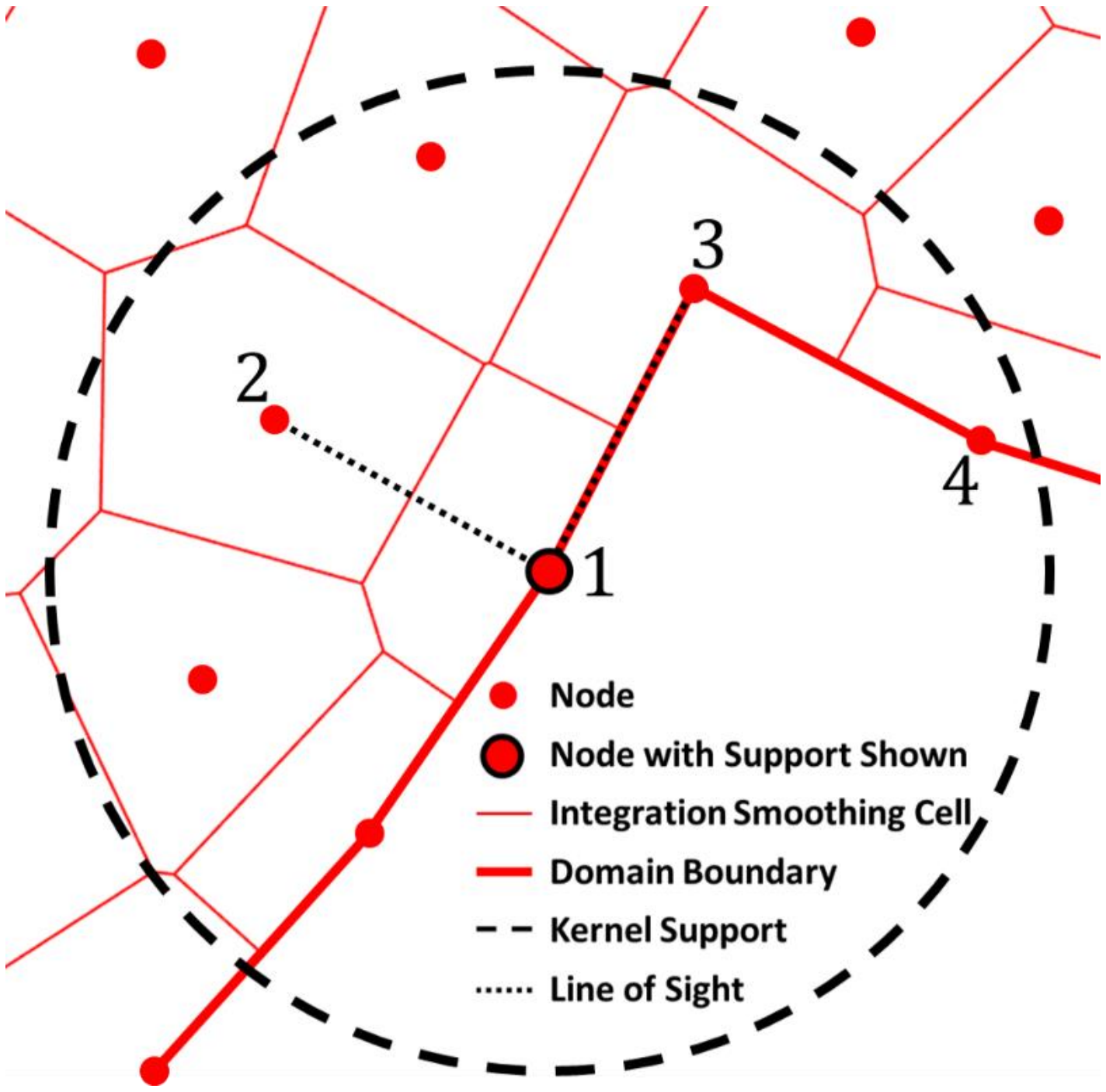

Figure 3: Example of line-of-sight

### 3.2. Domain Integration

#### 3.2.1. Stabilized Conforming/Non-conforming Nodal Integration

The Gauss quadrature rules commonly used in the finite element method are not suitable for meshfree methods due to the absence of a mesh in a meshfree discretization. While “background” meshes, independent of the discretization points, have been used to define Gauss quadratures, it is almost impossible to match the mesh with the supports of the RK kernel functions as shown in Figure 3, leading to integration errors and hence reduced convergence rates in meshfree methods [64, 65, 57, 59]. Higher-order quadratures have also been considered. For example, use of the cubic B-spline kernel in (33) in combination with a linear basis would require at least a 5$^{\text{th}}$ order Gauss quadrature rule for sufficient accuracy in common practice, which is computationally intractable [57, 59, 66, 67]. Alternatively, direct nodal integration (DNI) which locates an integration point at the node is natural to meshfree methods, but it under-integrates the strain energy and yields spurious zero energy modes [65].

These problems with conventional integration methods led to the development of stabilized conforming nodal integration (SCNI) [65], where a smoothed gradient is calculated over a conforming nodal “integration smoothing cell” rather than taking a direct derivative at the nodal integration points, see Figure 4. The following smoothed gradient has been introduced [65]:

$$\tilde{\Psi}_{I,i}(\boldsymbol{x}_L) = \frac{1}{V_L} \int_{\Omega_L} \Psi_{I,i}(\boldsymbol{x})\, d\Omega = \frac{1}{V_L} \int_{\partial\Omega_L} \Psi_I(\boldsymbol{x})\, n_i(\boldsymbol{x})\, d\Gamma, \tag{37}$$

where $\tilde{\Psi}_{I,i}(\boldsymbol{x}_L)$ is the smoothed shape function gradient in the $i^{\text{th}}$ direction of node $I$ which is constant over the cell of node $L$ located at $\boldsymbol{x}_L$, $V_L$ is the nodal cell volume of the integration smoothing cell $\Omega_L$ associated with node $L$, and $n_i$ is the $i^{\text{th}}$ component of the outward normal vector on the integration smoothing cell boundary $\partial\Omega_L$; see Figure 4. It has been shown that SCNI fulfills the integration constraint and meets linear exactness in the Galerkin approximation [65, 57]. The conforming integration smoothing cells may be constructed in several ways including Voronoi diagrams, Delaunay triangulations, or from a finite element mesh with the requirement that the cells' boundaries conform to each other.

To model extreme deformation and fragmentation problems, a non-conforming counterpart to SCNI was developed, the stabilized non-conforming nodal integration (SNNI) [68, 69]. In SNNI, the integration smoothing cells are constructed with simple geometries, such as cubes or spheres centered on the nodes, and whose volume is equal to that of their conforming counterparts. Two-dimensional examples of SCNI and SNNI cells are shown in Figure 4.

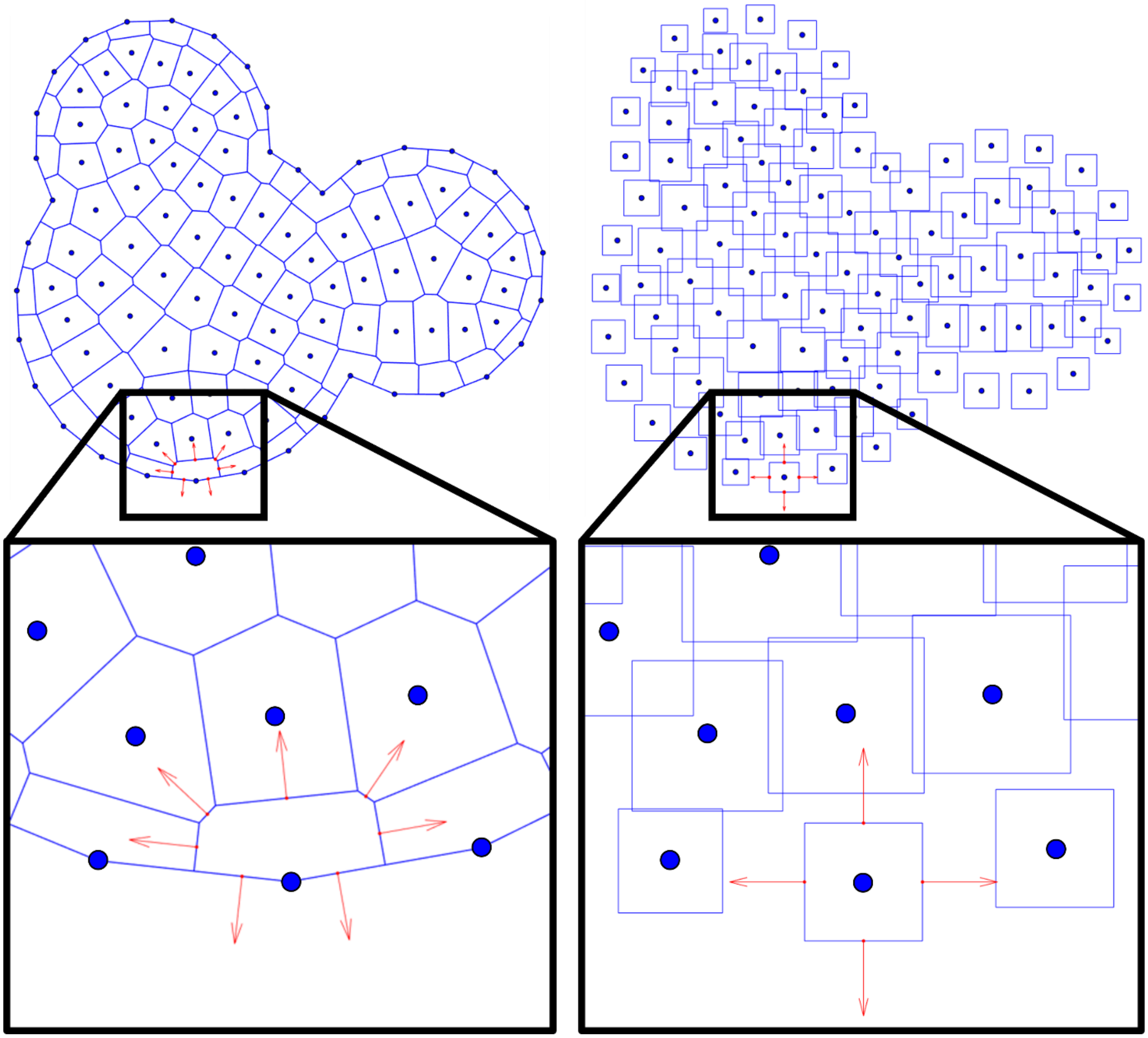

Figure 4: Smoothing integration cells; SCNI using a Voronoi diagram to achieve conformity (left) and SNNI using square non-conforming cells (right); red arrows indicate the outward normal vectors of the cells at the cell boundary evaluation points

### 3.2.2. Variationally Consistent Correction for Non-conforming Nodal Integration

For the proposed quasi-conforming embedded RKPM formulation, a SNNI will be employed which is essential to the effectiveness and reliability of the proposed method. Since the integration smoothing cells in SNNI no longer conform to each other, SNNI loses first order variational consistency and, consequently, optimal convergence rates. To restore variational consistency, a variationally consistent correction of integration methods that are not variationally consistent, including SNNI, was developed and termed variationally consistent SNNI (VC-SNNI) [66]. The correction allows a nodal integration with conforming or non-conforming integration smoothing cells to achieve variational consistency for any basis order.

Variational consistency is the ability of a Galerkin method to achieve $n^{\text{th}}$ order Galerkin exactness for optimal convergence. Variational consistency of the $n^{\text{th}}$ order requires $n^{\text{th}}$ order completeness, which can be achieved by RK with $n^{\text{th}}$ order bases, and an $n^{\text{th}}$ order integration constraint of the following form [57]:

$$\hat{\int_{\Omega}} \bar{\Psi}_{I,j}\, \boldsymbol{P}\, d\Omega = \hat{\int_{\partial\Omega}} \bar{\Psi}_I\, n_j\, \boldsymbol{P}\, d\Gamma - \hat{\int_{\Omega}} \bar{\Psi}_I\, \boldsymbol{P}_{,j}\, d\Omega, \tag{38}$$

where $\bar{\Psi}_I$ is the approximation function for the test function, "^" in $\hat{\int}$ denotes numerical integration, and $\boldsymbol{P}$ is the $(n\text{-}1)^{\text{th}}$ order monomial vector:

$$\boldsymbol{P} = \begin{bmatrix} 1 & x & y & x^2 & xy & y^2 & \cdots & x^{\alpha}y^{\beta} & \cdots & y^{(n-1)} \end{bmatrix}^T. \tag{39}$$

Eqn. (38) is called the integration constraint in [66]. For any Galerkin approximation to achieve a desired level of variational consistency, it must 1) have a trial function that can reproduce the desired basis order exactly in (36) and 2) satisfy the integration constraint in (38). For linear variational consistency, $\boldsymbol{P} = 1$, and Eqn. (38) reduces to:

$$\hat{\int_{\Omega}} \bar{\Psi}_{I,j}\, d\Omega = \hat{\int_{\partial\Omega}} \bar{\Psi}_I\, n_j\, d\Gamma, \tag{40}$$

which can be achieved by Eqn. (37) with conforming integration smoothing cells. This indicates that SCNI with a linear basis automatically meets first-order variational consistency and does not need VC correction. If we want a higher-order variational consistency in (38) for integration methods that are not variationally consistent, a modification to the test function gradient can be considered [66]:

$$\bar{\Psi}_{I,j} = \Psi_{I,j} + \Theta_I \boldsymbol{P}^T \boldsymbol{\zeta}_{Ij}, \tag{41}$$

where $\bar{\Psi}_{I,j}$ is the modified test function gradient, $\Theta_I$ is:

$$\Theta_I = \begin{cases} 1, & \operatorname{supp}(\Psi_I) \neq 0 \\ 0, & \text{else} \end{cases}, \tag{42}$$

and $\boldsymbol{\zeta}_{Ij}$ is a column vector of unknown coefficients for each node $I$ and direction $j$. Substituting the modified test function gradient in (41) in place of the standard test function gradient of (38) and rearranging:

$$\hat{\int_{\Omega}} \Theta_I \, \boldsymbol{P} \, \boldsymbol{P}^T d\Omega \; \boldsymbol{\zeta}_{Ij} = \hat{\int_{\partial\Omega}} \Psi_I \, n_j \, \boldsymbol{P} \, d\Gamma - \hat{\int_{\Omega}} \left( \Psi_I \, \boldsymbol{P}_{,j} + \Psi_{I,j} \, \boldsymbol{P} \right) d\Omega, \tag{43}$$

or:

$$\boldsymbol{M}_I \boldsymbol{\zeta}_{Ij} = \boldsymbol{r}_I^j, \tag{44}$$

$$\boldsymbol{M}_I = \int_{\Omega} \Theta_I \, \boldsymbol{P} \, \boldsymbol{P}^T d\Omega \,, \tag{45}$$

$$\boldsymbol{r}_I^j = \hat{\int_{\partial\Omega}} \Psi_I \, n_j \, \boldsymbol{P} \, d\Gamma - \hat{\int_{\Omega}} \left( \Psi_I \, \boldsymbol{P}_{,j} + \Psi_{I,j} \, \boldsymbol{P} \right) d\Omega \,, \tag{46}$$

where $\boldsymbol{r}_I^j$ is the residual of the formulation without VC correction (38). Once (43) is solved for $\boldsymbol{\zeta}_{Ij}$, the modified test function gradients (41) can be computed. Using the modified test function gradients, the Galerkin approximation of elasticity can be expressed as:

$$\hat{\int_{\Omega}} \bar{\boldsymbol{B}}_I^T \,\boldsymbol{\sigma}\, d\Omega = \hat{\int_{\partial\Omega}} \boldsymbol{\Psi}_I^T \,\boldsymbol{\eta}^T \,\boldsymbol{\sigma}\, d\Gamma + \hat{\int_{\Omega}} \boldsymbol{\Psi}_I^T \,\boldsymbol{b}\, d\Omega, \tag{47}$$

where $\bar{\boldsymbol{B}}_I$ is the modified strain-displacement matrix for node $I$ and is comprised of the modified test function gradients calculated from Eqn. (41). Substituting $\boldsymbol{\sigma} = \boldsymbol{C}\,\boldsymbol{B}\,\boldsymbol{d}$ into the LHS of (47), the stiffness matrix is:

$$\boldsymbol{K}_{IJ} = \hat{\int_{\Omega}} \bar{\boldsymbol{B}}_I^T \,\boldsymbol{C}\,\boldsymbol{B}_J\, d\Omega, \tag{48}$$

which is a Petrov-Galerkin formulation. It was shown in [57] that such a form is stable if the coefficients in $\boldsymbol{\zeta}_{Ij}$ are sufficiently small.

For the examples in this paper, we use a linear basis and seek to achieve first order variational consistency. For a linear basis, $\boldsymbol{P} = 1$, and equations (44) to (46) become:

$$M_I \zeta_{Ij} = r_I^j, \tag{49}$$

$$M_I = \hat{\int_{\Omega}} \Theta_I \, d\Omega\,, \tag{50}$$

$$r_I^j = \hat{\int_{\partial\Omega}} \Psi_I \, n_j \, d\Gamma - \hat{\int_{\Omega}} \Psi_{I,j} \, d\Omega\,. \tag{51}$$

### 3.2.3. Naturally Stabilized Nodal Integration

Although SCNI and VC-SNNI with linear bases achieve first-order variational consistency, they are still nodal integration methods and, consequently, reduced-order integration methods. As such, these integration methods can still trigger non-physical, low-energy modes which necessitate the use of stabilization methods as a remedy [70, 71, 66]. The most computationally efficient stabilization method that avoids tunable parameters is the naturally stabilized nodal integration (NSNI) [66]. NSNI expands the strain-displacement matrix in a Taylor series [72, 73, 74]:

$$\widehat{\boldsymbol{B}}_I(\mathbf{x}_L) = \widetilde{\boldsymbol{B}}_I(\mathbf{x}_L) + (x - x_L) \cdot \widetilde{\boldsymbol{B}}_{Ix}^{\nabla}(\mathbf{x}_L) + (y - y_L) \cdot \widetilde{\boldsymbol{B}}_{Iy}^{\nabla}(\mathbf{x}_L) + (z - z_L) \cdot \widetilde{\boldsymbol{B}}_{Iz}^{\nabla}(\mathbf{x}_L), \quad (52)$$

where $\widetilde{\boldsymbol{B}}_I$ is the smoothed shape function gradient matrix from SCNI or VC-SNNI following (37) and $\widetilde{\boldsymbol{B}}_{Ix}^{\nabla}$, $\widetilde{\boldsymbol{B}}_{Iy}^{\nabla}$, and $\widetilde{\boldsymbol{B}}_{Iz}^{\nabla}$ are the smoothed gradients of the implicit gradients in the $x$-, $y$-, and $z$-directions, respectively:

$$\widetilde{\boldsymbol{B}}_{Ix}^{\nabla} = \begin{bmatrix} \widetilde{\Psi}_{I,xx}^{\nabla} & 0 & 0 \\ 0 & \widetilde{\Psi}_{I,yx}^{\nabla} & 0 \\ 0 & 0 & \widetilde{\Psi}_{I,zx}^{\nabla} \\ \widetilde{\Psi}_{I,yx}^{\nabla} & \widetilde{\Psi}_{I,xx}^{\nabla} & 0 \\ \widetilde{\Psi}_{I,zx}^{\nabla} & 0 & \widetilde{\Psi}_{I,xx}^{\nabla} \\ 0 & \widetilde{\Psi}_{I,zx}^{\nabla} & \widetilde{\Psi}_{I,yx}^{\nabla} \end{bmatrix}, \quad (53)$$

$$\widetilde{\boldsymbol{B}}_{Iy}^{\nabla} = \begin{bmatrix} \widetilde{\Psi}_{I,xy}^{\nabla} & 0 & 0 \\ 0 & \widetilde{\Psi}_{I,yy}^{\nabla} & 0 \\ 0 & 0 & \widetilde{\Psi}_{I,zy}^{\nabla} \\ \widetilde{\Psi}_{I,yy}^{\nabla} & \widetilde{\Psi}_{I,xy}^{\nabla} & 0 \\ \widetilde{\Psi}_{I,zy}^{\nabla} & 0 & \widetilde{\Psi}_{I,xy}^{\nabla} \\ 0 & \widetilde{\Psi}_{I,zy}^{\nabla} & \widetilde{\Psi}_{I,yy}^{\nabla} \end{bmatrix}, \quad (54)$$

$$\widetilde{\boldsymbol{B}}_{Iz}^{\nabla} = \begin{bmatrix} \widetilde{\Psi}_{I,xz}^{\nabla} & 0 & 0 \\ 0 & \widetilde{\Psi}_{I,yz}^{\nabla} & 0 \\ 0 & 0 & \widetilde{\Psi}_{I,zz}^{\nabla} \\ \widetilde{\Psi}_{I,yz}^{\nabla} & \widetilde{\Psi}_{I,xz}^{\nabla} & 0 \\ \widetilde{\Psi}_{I,zz}^{\nabla} & 0 & \widetilde{\Psi}_{I,xz}^{\nabla} \\ 0 & \widetilde{\Psi}_{I,zz}^{\nabla} & \widetilde{\Psi}_{I,yz}^{\nabla} \end{bmatrix}, \tag{55}$$

where $\widetilde{\Psi}_{I,ij}^{\nabla}$ denotes the smoothed gradient in the $i^{\text{th}}$-direction using the implicit gradient shape function in the $j^{\text{th}}$-direction. The implicit gradient shape function in the $j^{\text{th}}$-direction is given as:

$$\Psi_{I,j}^{\nabla}(\boldsymbol{x}) = \mathbf{H}_j^T \, \mathbf{M}^{-1}(\boldsymbol{x}) \, \mathbf{H}(\boldsymbol{x} - \boldsymbol{x}_I) \, \phi_a(\boldsymbol{x} - \boldsymbol{x}_I), \tag{56}$$

where $\mathbf{H}_j$ for a linear basis has the following values:

$$\mathbf{H}_x^T = [0, -1, 0, 0], \tag{57}$$
$$\mathbf{H}_y^T = [0, 0, -1, 0], \tag{58}$$
$$\mathbf{H}_z^T = [0, 0, 0, -1]. \tag{59}$$

Note that (56) has the same form as the standard RK shape function equation (32) with the only difference being the substitution of $\mathbf{H}_j^T$ in place of $\mathbf{H}^T(\mathbf{0})$. As a result, the NSNI stabilization requires only a minimal computational effort.

With NSNI in (52), the stiffness matrix with VC correction becomes:

$$\hat{\int_{\Omega}} \widehat{\widetilde{\boldsymbol{B}}}_I^T \boldsymbol{C} \, \widehat{\boldsymbol{B}}_J \, d\Omega = \sum_{L=1}^{NP} \left( \widetilde{\widetilde{\boldsymbol{B}}}_I^T \, \boldsymbol{C} \, \widetilde{\boldsymbol{B}}_J \, V_L + \widetilde{\boldsymbol{B}}_{Ix}^{\nabla\,T} \boldsymbol{C} \, \widetilde{\boldsymbol{B}}_{Jx}^{\nabla} \, M_{Lx} + \widetilde{\boldsymbol{B}}_{Iy}^{\nabla\,T} \boldsymbol{C} \, \widetilde{\boldsymbol{B}}_{Jy}^{\nabla} \, M_{Ly} + \widetilde{\boldsymbol{B}}_{Iz}^{\nabla\,T} \boldsymbol{C} \, \widetilde{\boldsymbol{B}}_{Jz}^{\nabla} \, M_{Lz} \right), \tag{60}$$

where $\widehat{\widetilde{\boldsymbol{B}}}_I$ is the VC-corrected, expanded strain-displacement matrix, $\widetilde{\widetilde{\boldsymbol{B}}}_I$ is the VC-corrected, smoothed strain-displacement matrix from SCNI or SNNI following (41) and (53) through (59), and $M_{Lx}$, $M_{Ly}$, and $M_{Lz}$ are the second moments of inertia about node $L$. In (60), we have assumed that the first moments of inertia and products of inertia are zero when nodes are located at the centroids of their nodal cells. While boundary nodes are not located at their cell centers, we have found from numerical experiments that the second moments of inertia are sufficient to stabilize these nodes.

As we will show in section 4, the background matrix domain will be comprised of conforming smoothing integration cells except near the interface where the cells are non-conforming, while the foreground inclusion domain(s) will be comprised exclusively of conforming cells. Thus, VC-SNNI and SCNI, both stabilized with NSNI, will be used for the background and foreground domains, respectively. Direct nodal integration will be used for all other volume integrals such as the body force terms.

## 4. Quasi-Conforming Embedded Discretization

The proposed quasi-conforming embedded (QCE) discretization is a three-stage process, and this section details the reasoning behind the geometric manipulations at each stage. The advantages of this meshfree strategy over standard immersed and embedded techniques rooted in mesh-based methods such as FEM will be noted throughout.

### 4.1. Initial Immersed State

To avoid time-consuming, conformal discretization, the background and foreground domains are discretized independently, and the foreground discretization is initially immersed in the background discretization as shown in Figure 5. The region shared by both background and foreground domains is called the *overlapping region*, and the background discretization in the overlapping region is called the *fictitious discretization*.

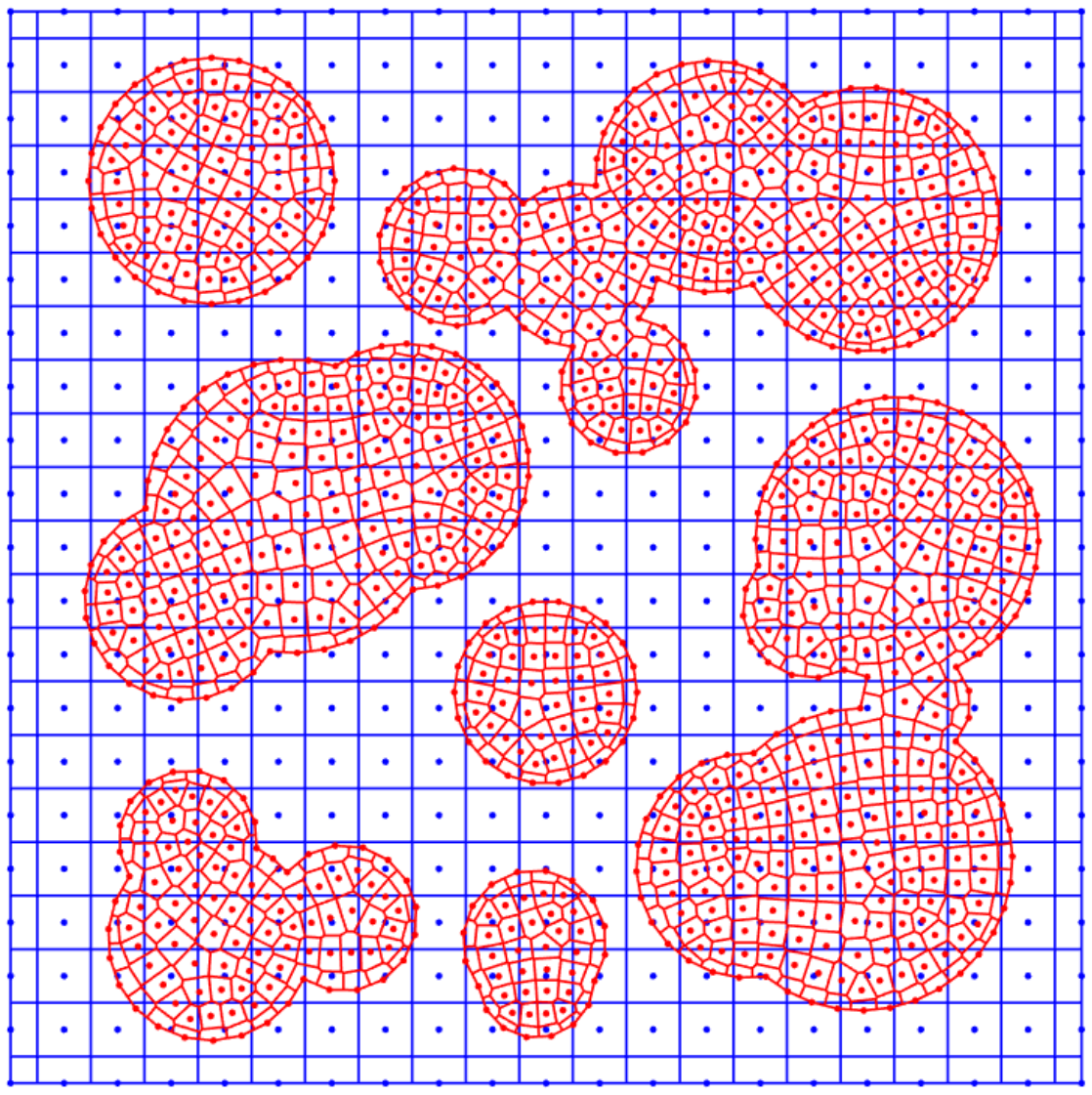

Figure 5: Background and foreground smoothing integration cells at the initial immersed state

### 4.2. Non-conforming Quadtree Subdivision

One problem with immersed methods (especially VCIMs) is the "smearing" of the background solution caused by that approximation's continuity across the interface. Capturing the weak discontinuity at the interface is essential when modeling, for example, the debonding of inclusions in composite materials. The second problem with immersed methods is that the discretization ratio (between background and foreground domains) must stay in a certain range to prevent the ill-conditioning of the stiffness matrix. In our proposed framework, we remove the fictitious domain nodes and cells thereby preventing the smearing of the background approximation and the ill-conditioning of the stiffness matrix. *For RKPM with RK approximation, these discretization cells are used as a "reference" for computing nodal areas (volumes) and as the integration smoothing cells for SCNI or SNNI strain smoothing.*

For the matrix integration points to fulfill the first order integration constraint in a *non-conforming manner*, the gap resulting from the complete removal of the fictitious smoothing integration cells that intersect the interface can be recovered easily by correcting the area (volume) of these cells. In the case where the background to foreground nodal spacing ratio is too large (and vice versa), node and smoothing integration cell insertions can be done systematically, see Figure 6.

A simple strategy to determine the level of sub-division of the background smoothing integration cells, denoted as $n_R$, is given as follows:

$$n_R = \begin{cases} \lfloor n_{log} \rfloor, & n_{log} > 0 \\ 0, & n_{log} \leq 0 \end{cases}, \tag{61}$$

$$n_{log} = \log_2(\mathcal{R}'), \tag{62}$$

$$\mathcal{R}' = \begin{cases} \lceil \mathcal{R} \rceil, & \mathcal{R} \bmod 1 \geq k,\ \mathcal{R} > 1 \\ \lfloor \mathcal{R} \rfloor, & \mathcal{R} \bmod 1 < k,\ \mathcal{R} > 1, \\ 1, & \mathcal{R} \leq 1 \end{cases} \tag{63}$$

$$\mathcal{R} = \frac{h^-}{h^+_{\text{Intf}}}, \tag{64}$$

where $n_R$ is the number of times background smoothing integration cells intersecting the interface are subdivided, $\lfloor \cdot \rfloor$ is the floor function which rounds a fraction down to the nearest integer, $\lceil \cdot \rceil$ is the ceiling function which rounds a fraction up to the nearest integer, $k$ is a real number between 0 and 1, $\mathcal{R}$ is the nodal spacing ratio, $h^-$ is the uniform background nodal spacing before quadtree subdivision, and $h^+_{\text{Intf}}$ is the averaged nodal spacing of the inclusion nodes on the interface. In our examples, as demonstrated in Figure 6, we have used $k = 0.5$ which simply makes (63) round $\mathcal{R}$ to the nearest integer unless $\mathcal{R} \leq 1$. This quadtree subdivision provides a smooth discretization density transition near the material interfaces. A similar strategy can be defined in the case where the background discretization density is higher than the foreground discretization density. Figure 7(a) and Figure 7(b) show one-level and two-level non-conforming quadtree subdivision intermediate states, respectively. This non-conforming quadtree subdivision strategy with variable subdivision levels can also be considered, as shown in Figure 7(c).

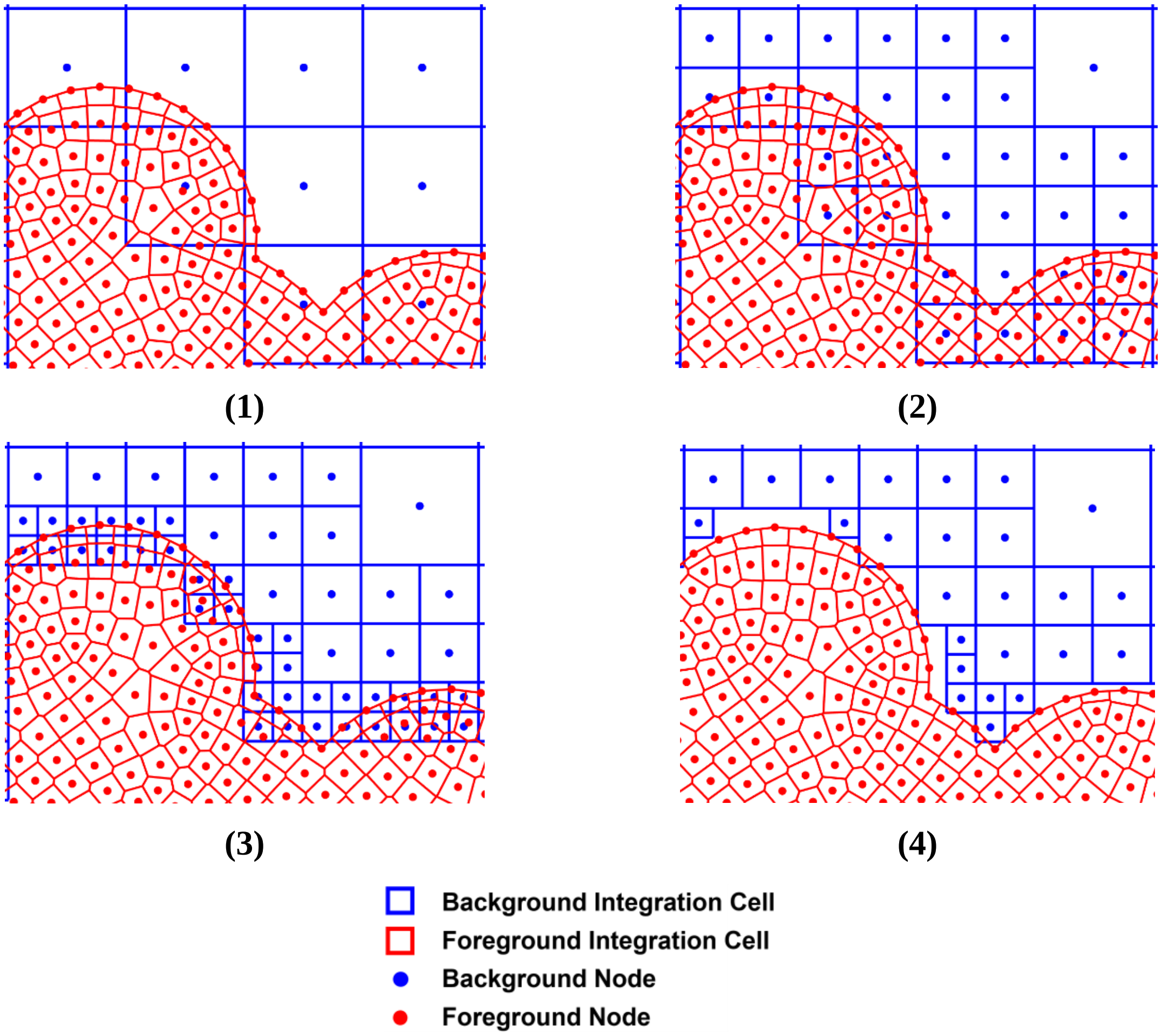

Figure 6: Quadtree subdivision process: (1) Remove all nodes and smoothing integration cells that are completely in the overlapping region. (2) If $n_R \geq 1$, refine remaining smoothing integration cells within a specified distance or that intersect the interface. (3) If $n_R \geq 2$, repeat steps (1) and (2) above $n_R - 1$ more times for cells that straddle the interface. (4) Remove any new smoothing integration cells with cell boundary evaluation points in the overlapping region.

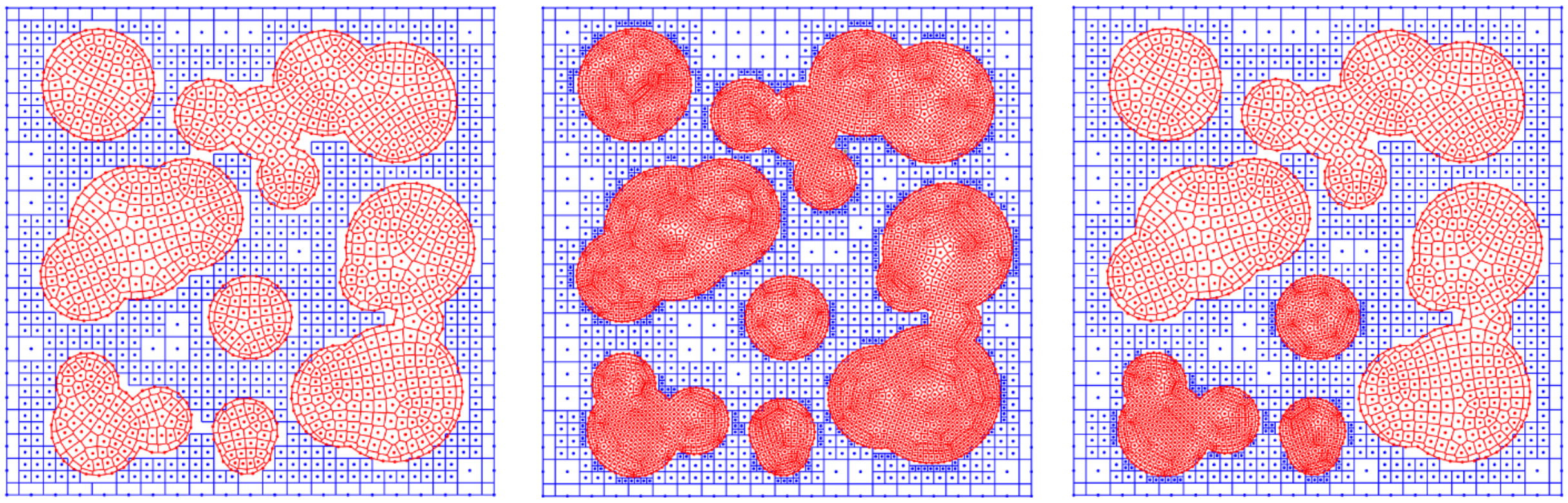

(a) (b) (c)

Figure 7: Non-conforming quadtree subdivisions: (a) one-level subdivision, (b) two-level subdivision, and (c) variable-level quadtree subdivisions

### 4.3. Integration Constraint Correction of Non-conforming Nodal Integration

In this section, we investigate the influence of the missing area (volume) due to the removal of overlapping smoothing cells and quadtree subdivisions. It is worth noting that the missing area (volume) can be corrected by the residual $\boldsymbol{r}_I^j$ in the VC correction of non-conforming nodal integration in (44) and (46). As mentioned in Section 3.2.2, the variationally consistent (VC) correction leads to a Petrov-Galerkin formulation, the stability of which depends on the magnitude of the residual $\boldsymbol{r}_I^j$ [57]. Thus, although it is not required that the volume of all smoothing cells equals the volume of the computational domain for the VC correction in (41) to work, the correction residual $\boldsymbol{r}_I^j$ is reduced by meeting the area (volume) conservation for better stability. The influence of the area (volume) conservation on the correction coefficients can be shown by replacing the RK shape functions in the first order integration constraint in (41) and (51) with a linear function. Let $\Psi_I = \Psi = x + y + z$ whose support covers the entirety of an arbitrary three-dimensional domain $\Omega$. Then from (51), we have:

$$r^j = \hat{\int}_{\partial\Omega} (x + y + z)\, n_j \, d\Gamma - \hat{\int}_{\Omega} d\Omega \,, \tag{65}$$

Using Gauss's theorem:

$$\int_{\partial\Omega} x_i \, n_j \, d\Gamma = \int_{\Omega} x_{i,j} \, d\Omega = \int_{\Omega} \delta_{ij} \, d\Omega = V\delta_{ij}. \tag{66}$$

where $V$ is the volume of the computational domain. Using (66) in (65), we have:

$$r^j = V - \hat{V}, \tag{67}$$

where $\hat{V}$ is the total volume of all integration smoothing cells:

$$\hat{\int}_{\Omega} d\Omega = \sum_{I=1}^{NC} V_I \equiv \hat{V}, \tag{68}$$

in which $NC$ is the number of smoothing cells and $V_I$ is the volume of each smoothing cell. Since the support of $\Psi_I = \Psi$ extends over the entire computational domain, then $\Theta_I = \Theta = 1$ for $\forall \boldsymbol{x} \in \Omega$, and since the integral in (50) is over all smoothing cells, (50) becomes:

$$M_I = M = \hat{\int}_{\Omega} \Theta \, d\Omega = \hat{\int}_{\Omega} d\Omega = \sum_{I=1}^{NC} V_I \equiv \hat{V}, \tag{69}$$

Therefore, (49) becomes:

$$\hat{V} \, \zeta_j = V - \hat{V}, \tag{70}$$

or:

$$\zeta_j = \frac{V - \hat{V}}{\hat{V}} = \frac{V}{\hat{V}} - 1, \tag{71}$$

Using (71) and noting that $\boldsymbol{P}^T = P = 1$ for a linear basis, (41) becomes:

$$\bar{\Psi}_{I,j} = \bar{\Psi}_{,j} = 1 + \frac{V}{\hat{V}} - 1 = \frac{V}{\hat{V}}. \tag{72}$$

Eqns. (71) and (72) show that the VC correction, in addition to meeting variational consistency, also acts as a volume correction, where the size of the VC correction is partially driven by the deviation of the total integration smoothing cell area (volume) from the computational domain area (volume).

**Remark 4.1**. *Enforcing area (volume) conservation in the integration smoothing cells, though non-conforming, allows the VC correction to contribute mainly to the correction of integration smoothing cell non-conformity, resulting in a more stable and accurate Galerkin solution. Integration smoothing cell volume conservation can be easily achieved by adding non-conforming "volume-recovery" smoothing cells centered on the interface nodes, as shown in Figure 8.*

**Remark 4.2.** *The interface nodes can be shared by both foreground and background domain approximations, which has the following advantages: 1) as discussed in Section 2.2, the enhanced coercivity allows for a penalty-free Nitsche's method, and 2) the partition of unity in the RK approximation can be maintained in each of the sub-domains.*

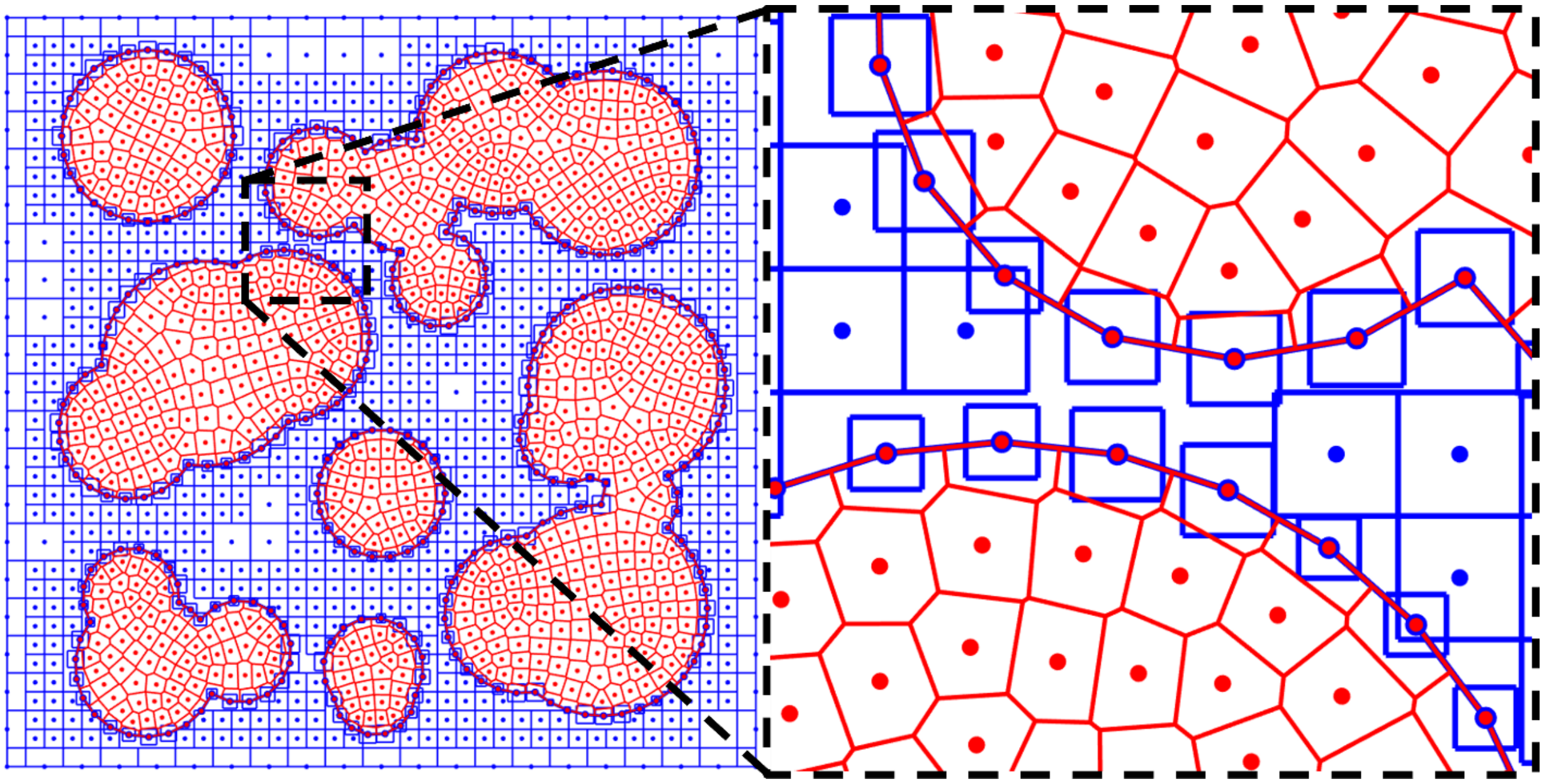

Figure 8: Addition of square, non-conforming integration smoothing cells centered on interface nodes

## 5. Numerical Examples

In this section, the effectiveness of the proposed QCE-RKPM is demonstrated through several numerical examples with comparisons to body-fitted FEM and analytical solutions. For all examples, the RK approximation uses a linear basis and a cubic B-spline kernel function with a normalized support size of $c = 2.0$, and the line-of-sight algorithm (see Section 3.1 and the modification in Section 4.3) is used to determine the influence of nodes. Additionally, the interface nodes are shared by both foreground and background domain approximations. SCNI and VC-SNNI are utilized for the domain integration of the foreground and background domains, respectively, and NSNI is employed for the stabilization of both. Nitsche's method with a penalty parameter of $\beta = 100\frac{E^-}{h^-}$ is utilized for enforcing Dirichlet boundary conditions where $h^-$ is the original nodal spacing of the background domain prior to quadtree subdivision. The quadtree subdivision of the background integration smoothing cells near the interface follows the strategies given in Section 4.2. For post-processing the QCE-RKPM results, the nodal smoothed gradients are used to calculate strain values at the nodes for those nodes with integration smoothing cells, while the direct gradient evaluated at the node is used for those nodes without integration smoothing cells. The RK shape functions are then utilized to interpolate those values to the Gauss points which have the dual purpose of both error calculation and contour plotting. To make a fair comparison, the FEM strains are calculated using the standard shape function gradients evaluated at the nodes for each element, and the nodal strains are averaged over all elements to which a node belongs. The element shape functions are then used to interpolate averaged nodal strain values to the Gauss points. To calculate error norms, $10^{th}$ order Gauss quadratures are used for sufficient accuracy.

### 5.1. One-dimensional composite bar

This example is used to examine how the "non-conforming" integration smoothing cells in the proposed QCE-RKPM affect the solution accuracy and convergence rates. As shown in Figure 9, a fixed-end composite bar is subjected to different loading and boundary conditions in the following studies. The linear elastic material properties are set to be $E^+ = 100{,}000$ and $E^- = 1{,}000$.

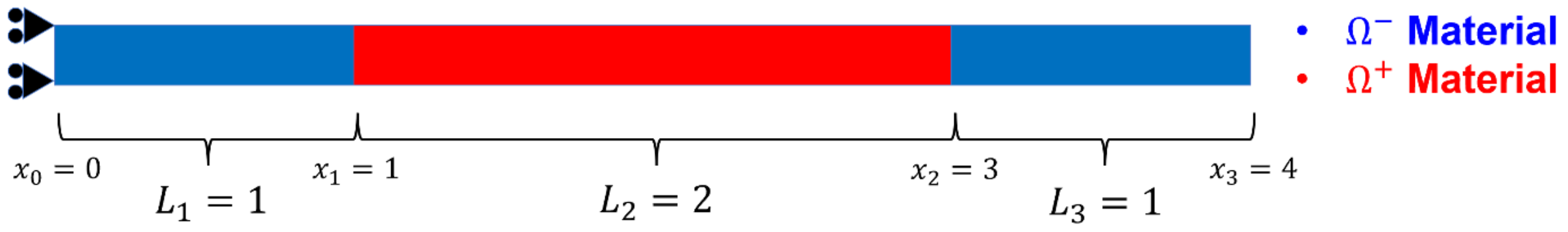

Figure 9: Composite bar with left end fixed

### 5.1.1. A piece-wise linear patch test with weak discontinuities

In this sub-section, a displacement of $u = g = 0.3$ is prescribed at the right end of the composite bar. As shown in Figure 10, the foreground domain $\Omega^+$ is discretized with 30 nodes and is initially immersed in the entire domain $\overline{\Omega} = \overline{\Omega}^- \cup \overline{\Omega}^+$ which is also discretized by 30 nodes. This creates the situation where the foreground nodal spacing is half that of the background nodal spacing and such that their corresponding integration smoothing cells do not match at the interface as shown in Figure 10.

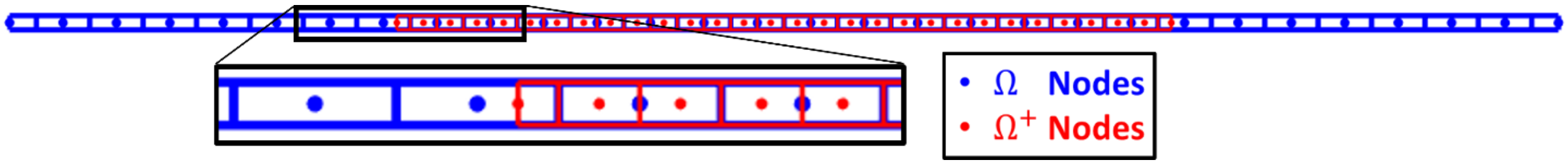

Figure 10: Pre-QCE immersed discretization

As outlined in Section 4.2, the background nodes and integration smoothing cells completely inside the foreground discretization are deleted, and background cells within $1.5h^+$ of the interface are divided in two. Any background cells still intersecting the interface are removed. Two cases were created to test the effect of the volume-recovery integration smoothing cells discussed in Section 4.3, one with these cells present (Figure 11(a)) and one without (Figure 11(b)). Lastly, the foreground interface nodes are shared with the background discretization.

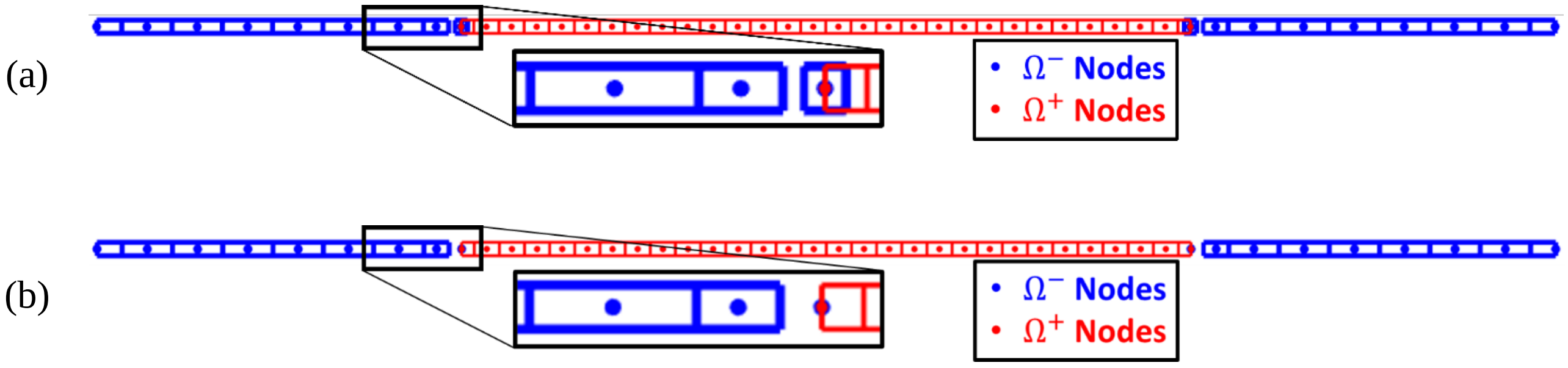

Figure 11: QCE-RKPM discretization for the 1D bar problem: (a) volume-recovery cells added around interface nodes and (b) no volume-recovery cells

As expected, both QCE-RKPM cases exactly reproduce the analytical solution at the nodes to nearly machine precision and give $L_2$ error norms of $1.6 \times 10^{-13}$ for the case with volume-recovery cells and $1.8 \times 10^{-13}$ for the case without these cells. The ability of QCE-RKPM to obtain an exact solution in this manufactured problem is due to the VC-SNNI that meets first order variational consistency *even with non-conforming integration smoothing cells*. Figure 12 shows the displacement, strain, and stress plots.

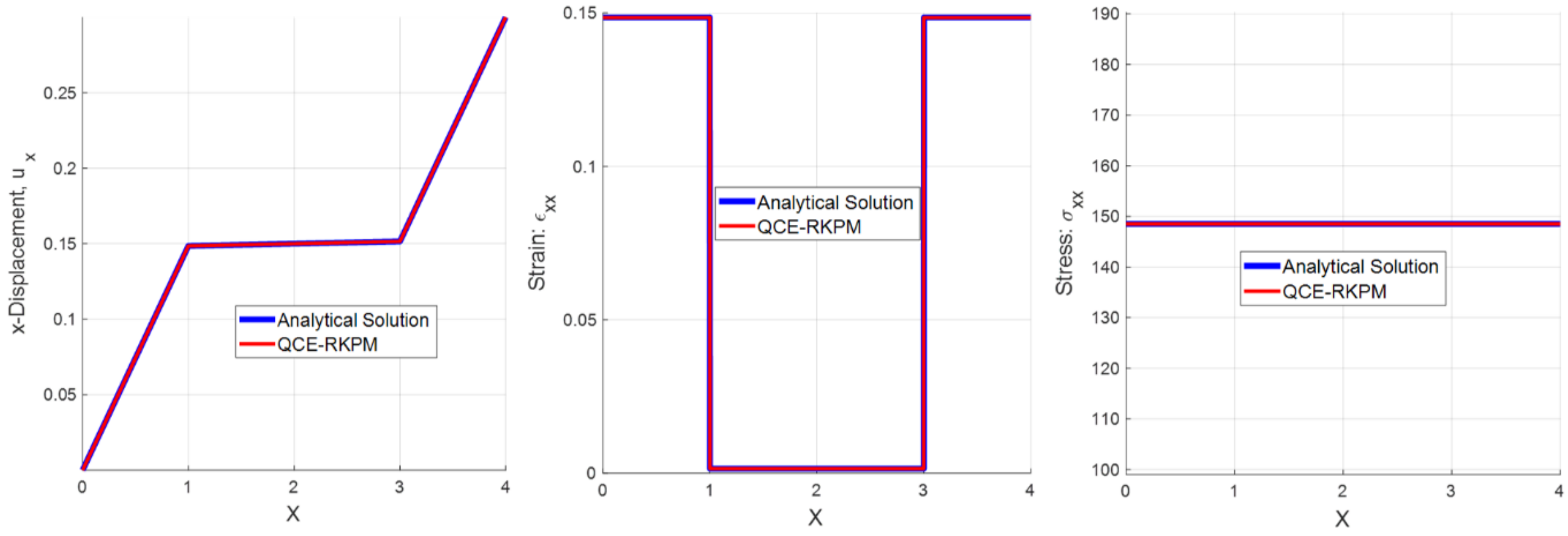

Figure 12: Displacement (left), strain (center), and stress (right) plots for both QCE-RKPM cases

### 5.1.2. Piecewise higher-order solution with weak discontinuities

This problem exhibits a piecewise higher-order solution and is designed to examine QCE-RKPM's convergence properties when linear basis functions are used. The composite bar is subjected to piecewise sinusoidal body force loadings as follows:

$$b^{-} = 10\sin\left(\frac{\pi x}{L_1}\right), \qquad 0 \le x \le x_1, \tag{73}$$

$$b^{+} = 50\sin\left(\frac{\pi(x - x_1)}{L_2}\right), \qquad x_1 \le x \le x_2, \tag{74}$$

$$b^{-} = 10\sin\left(\frac{\pi(x - x_2)}{L_3}\right), \qquad x_2 \le x \le x_3. \tag{75}$$

The body force terms (see Eqns. (85) and (91) in the Appendix) are integrated with direct nodal integration.

Figure 13 shows the QCE-RKPM displacement, strain, and stress results for the discretizations shown in Figure 11 along with the analytical solution. The results for the QCE-RKPM case with volume recovery cells (Figure 13(a)) are shown to be highly accurate compared with the exact solution, and while the QCE-RKPM case with no volume-recovery cells (Figure 13(b)) does show the expected inaccuracy near the material interfaces, the results are still very good.

(a)

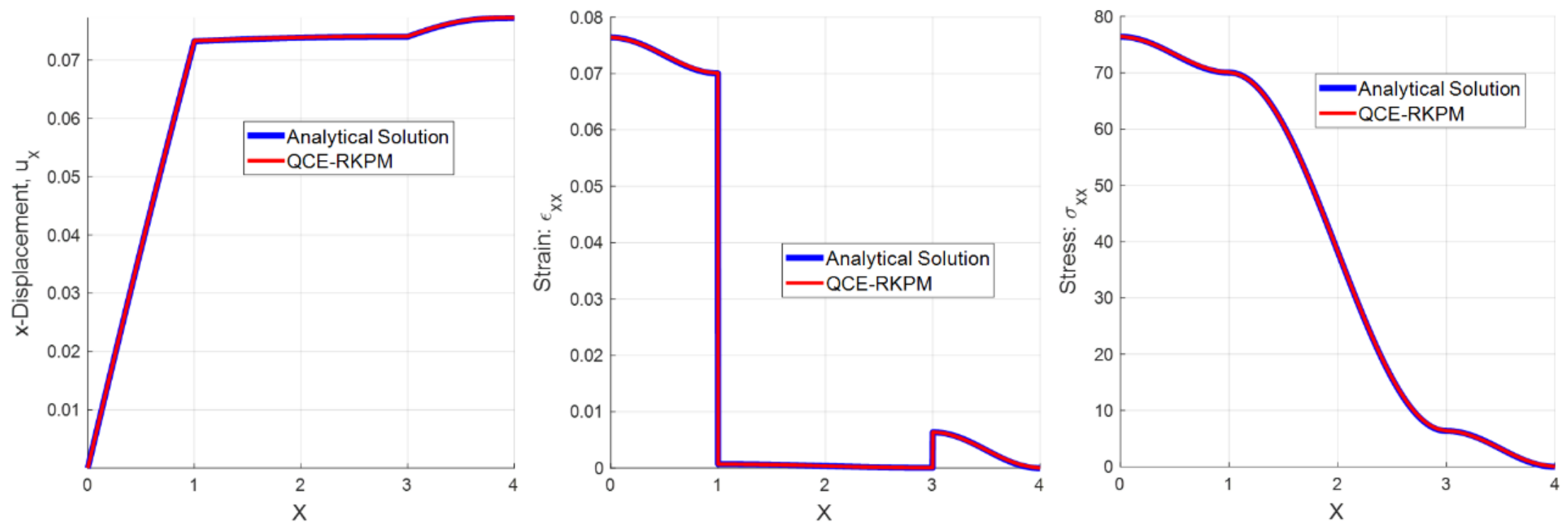

(b)

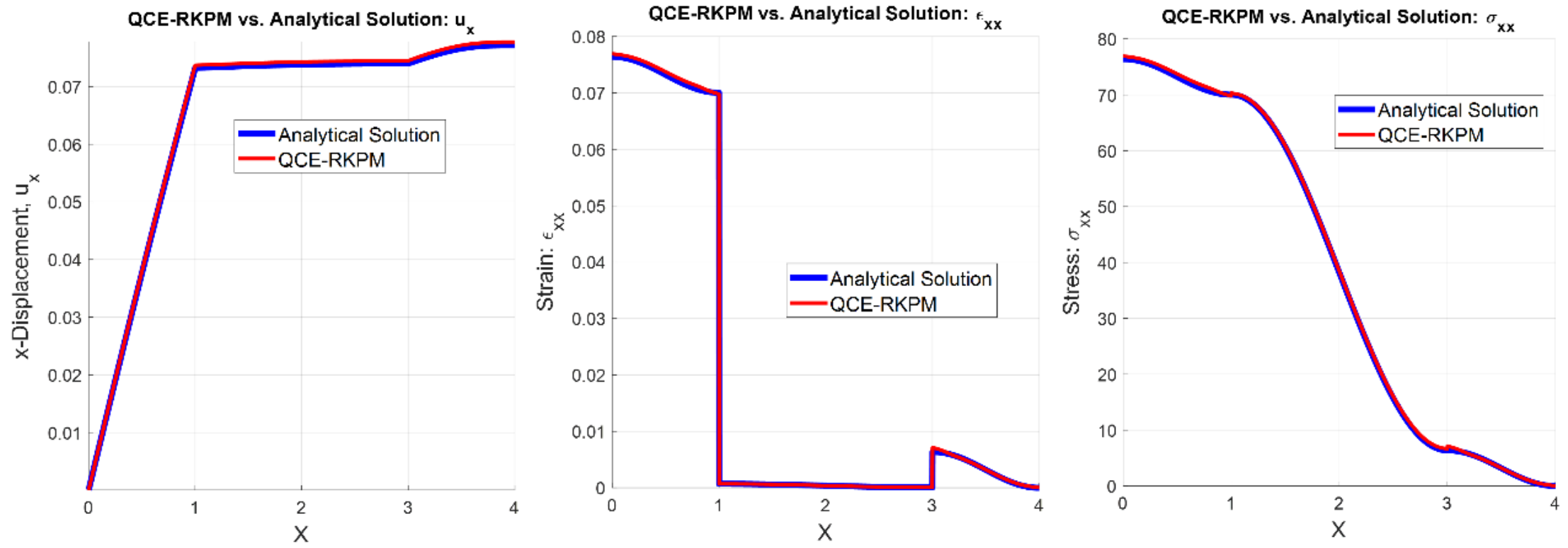

Figure 13: Displacement (left), strain (center), and stress (right) plots: (a) QCE-RKPM case with volume-recovery cells and (b) QCE-RKPM case without these cells

Figure 14 shows the convergence plots in the $L_2$ norm and $H_1$ semi-norm for both QCE-RKPM cases as well as body-fitted FEM. The numbers in the legends are the convergence rates. As shown in Figure 14, both QCE-RKPM cases converge at an optimal rate in the $L_2$ norm and possess super-convergence in the $H_1$ error semi-norm due to the re-interpolation of the smoothed and averaged nodal strain values, respectively. While the QCE-RKPM overall error is higher in the $L_2$ norm compared to FEM, and the QCE-RKPM with volume-recovery cells is shown to have roughly the same error level as body-fitted FEM.

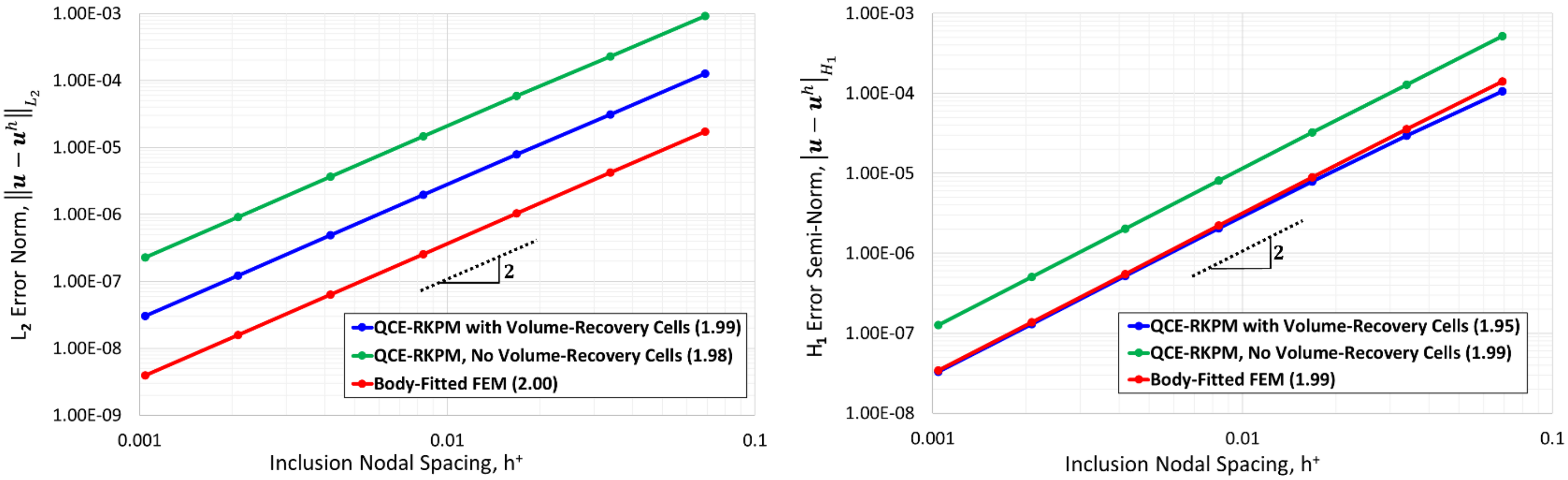

Figure 14: Convergence plots for the $L_2$ error norm (left) and $H_1$ error semi-norm (right)

## 5.2. Circular inclusion in an infinite plate subjected to far-field traction

In this example, an infinite plate with a circular inclusion of diameter $D = 2$ is subjected to a far-field traction $F = 100$ as shown in Figure 15.

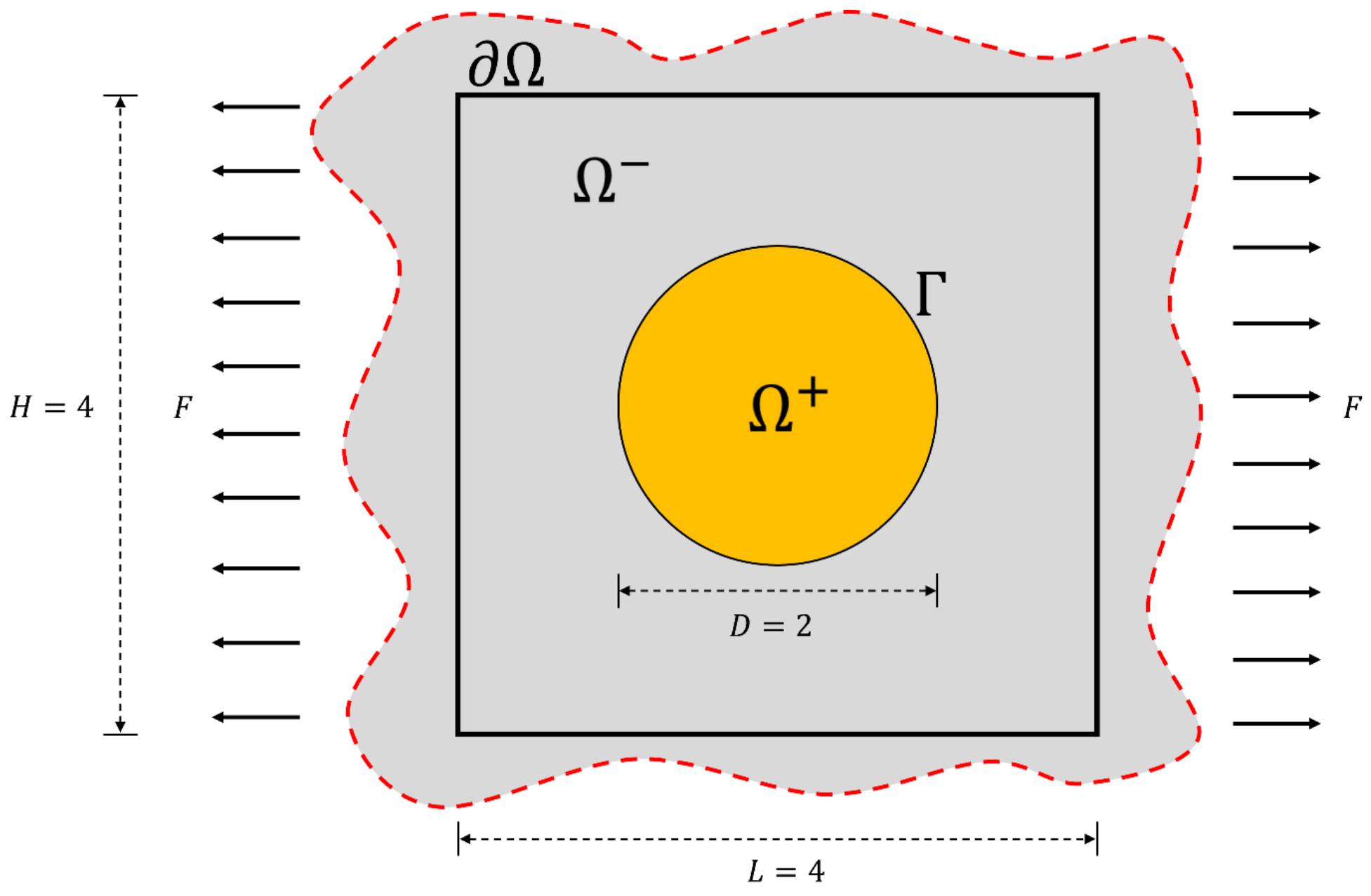

Figure 15: Infinite plate with circular inclusion subjected to far-field traction

The computational domain is a $4 \times 4$ square with outer boundary $\partial\Omega$ and an interface $\Gamma$ between the background matrix domain and foreground inclusion domain as shown in Figure 16. The elastic moduli used for this example are $E^{+} = 100{,}000$ and $E^{-} = 1{,}000$, and the Poisson's ratios are $\nu^{+} = \nu^{-} = 0.3$. The problem is modeled as plane stress. This problem possesses an analytical solution which is applied as a Dirichlet boundary condition to the entire background outer boundary $\partial\Omega$. The analytical solution to this problem can be found in [75]. Three discretization cases were run for the convergence study, two QCE-RKPM cases and one body-fitted FEM case. The two QCE-RKPM cases use an initial background nodal spacing which is approximately twice the spacing of the foreground interface nodes with quadtree subdivision around the interface. The first QCE-RKPM case uses square, volume-recovery non-conforming integration smoothing cells centered on the interface nodes (Figure 16(a) and Figure 17(a)), while the second QCE-RKPM case does not employ these volume-recovery integration smoothing cells (Figure 16(b) and Figure 17(b)). Note that, for both QCE-RKPM cases shown in Figure 17, the foreground interface nodes are shared with the background domain for enhanced coercivity in a penalty-free Nitsche's method as well as ensuring the partition of unity in both domains, as discussed in Sections 2.2 and 4.3. The corresponding body-fitted FEM case has a gradual element refinement analogous to the quadtree refinement as shown in Figure 16(c) for comparison.

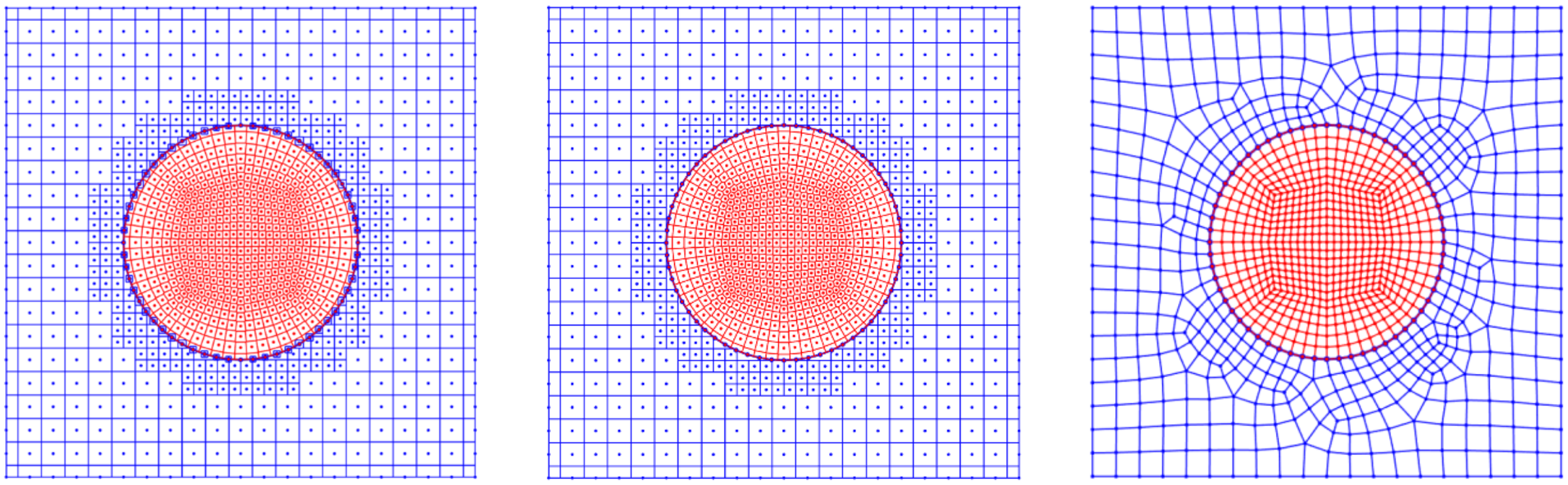

(a) (b) (c)

Figure 16: $h^{-} = 0.2$, $h_{\mathrm{Intf}}^{+} \approx 0.1$: (a) QCE-RKPM with volume-recovery integration smoothing cells centered on interface nodes, (b) QCE-RKPM without the volume-recovery smoothing cells, (c) and body-fitted FEM with gradual refinement

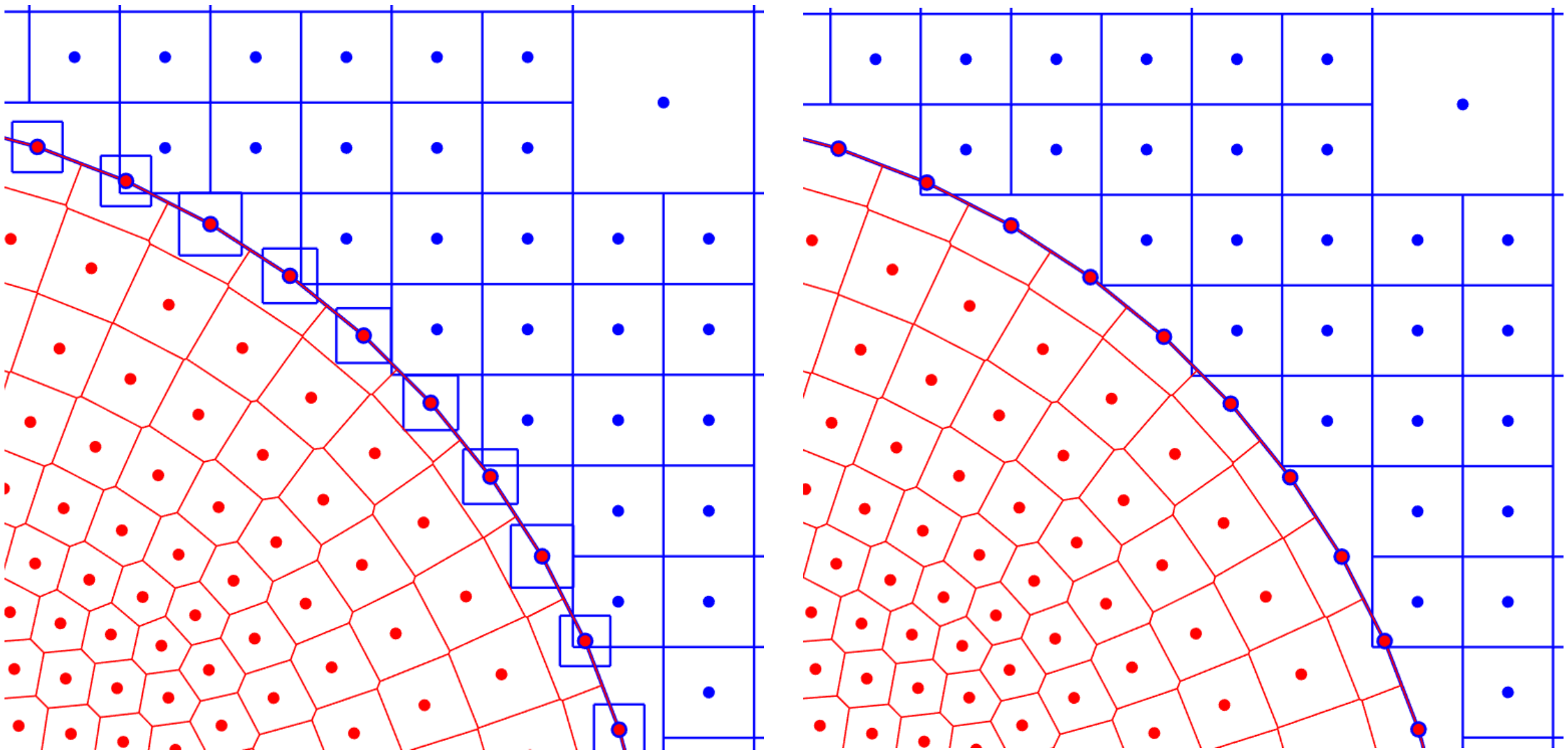

(a) (b)

Figure 17: Magnified views of (a) Figure 16(a) and (b) Figure 16(b)

Figure 18 shows the strain fields and the $\sigma_{yy}$ field for the discretizations in Figure 16 as well as the analytical solution. The contour plots for the QCE-RKPM case with volume-recovery integration smoothing cells show very little difference between them and the analytical solution, demonstrating the effectiveness of the proposed QCE-RKPM in achieving an accurate solution without the tedious effort in constructing a body-fitted discretization. The contour plots for the QCE-RKPM case without volume-recovery cells, while showing good results in the strain fields, do show a deviation in the stress from the analytical solution inside the inclusion region. This shows that conserving area (volume) with volume-recovery cells allows for more integration points and for the VC correction to contribute mainly to the lack of variational consistency thereby increasing the accuracy of the approximation. The body-fitted FEM solution, while superior to the QCE-RKPM case without volume-recovery cells, appears rough and asymmetric in places and is inferior in most respects to the QCE-RKPM case with volume-recovery integration smoothing cells.

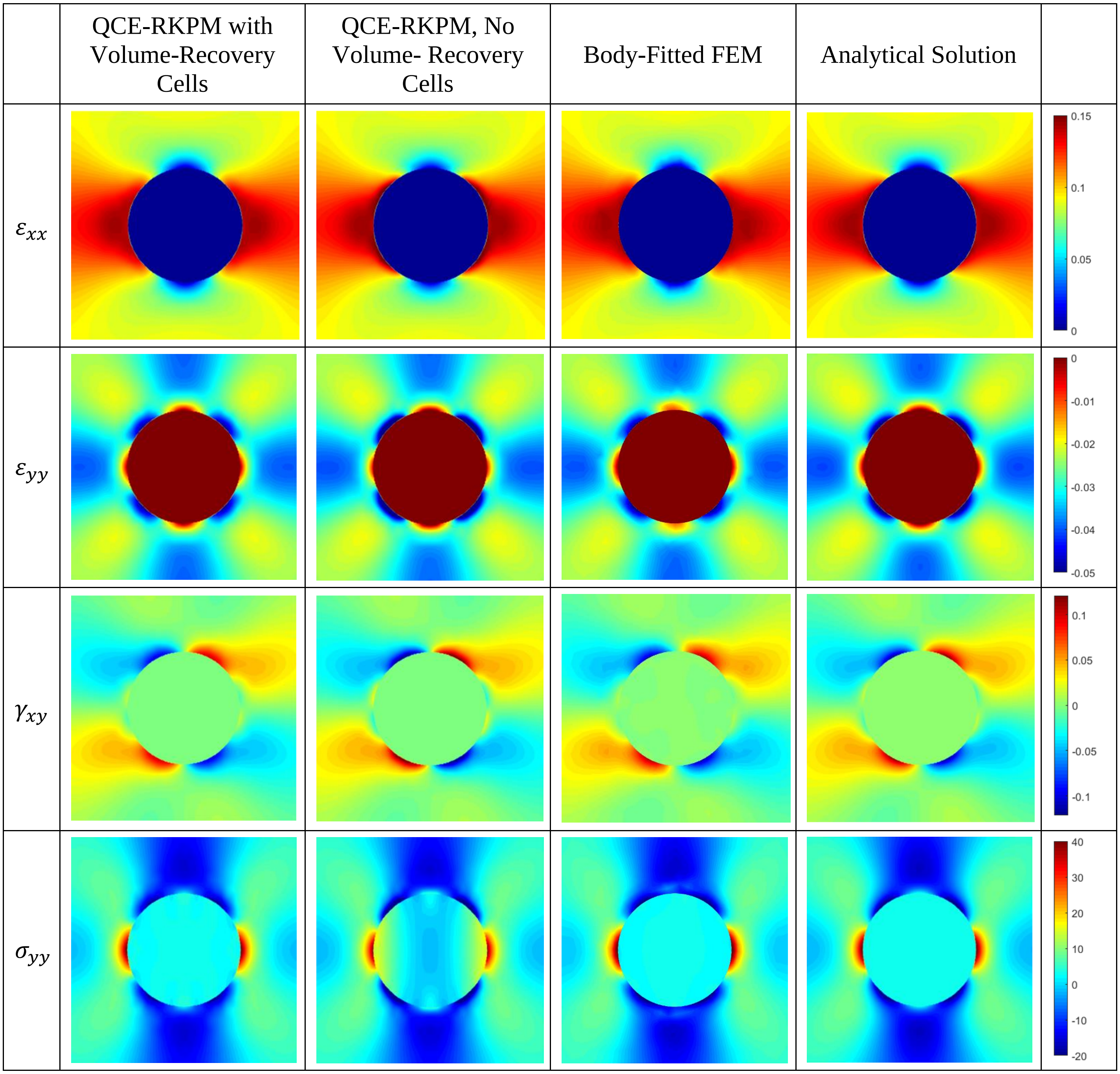

Figure 18: $\varepsilon_{xx}$, $\varepsilon_{yy}$, $\gamma_{xy}$, and $\sigma_{yy}$ fields for the discretizations in Figure 16 and the analytical solution

Figure 19 plots cross-sections of the normal strains and stresses along $y = 0$ for the discretizations in Figure 16. Again, these cross-sections show that the QCE-RKPM formulation using volume-recovery cells, though non-conforming, provides better accuracy than body-fitted FEM without the tedium of creating a body-fitted mesh. Cross-sections of the normal strains and stresses for the most refined case ($h^{+}_{\text{Intf}} \approx h^{-} = 0.0125$) are plotted in Figure 20 and show that both QCE-RKPM cases converge to the analytical solution with refinement. The convergence studies are shown in the plots of Figure 21. These plots demonstrate that QCE-RKPM with volume-recovery cells, while non-conforming, can achieve optimal convergence properties and similar overall error levels in the $L_2$ error norm as those of body-fitted FEM, and both QCE-RKPM cases achieve a super-convergent rate and superior overall error levels in the $H_1$ error semi-norm compared to body-fitted FEM.

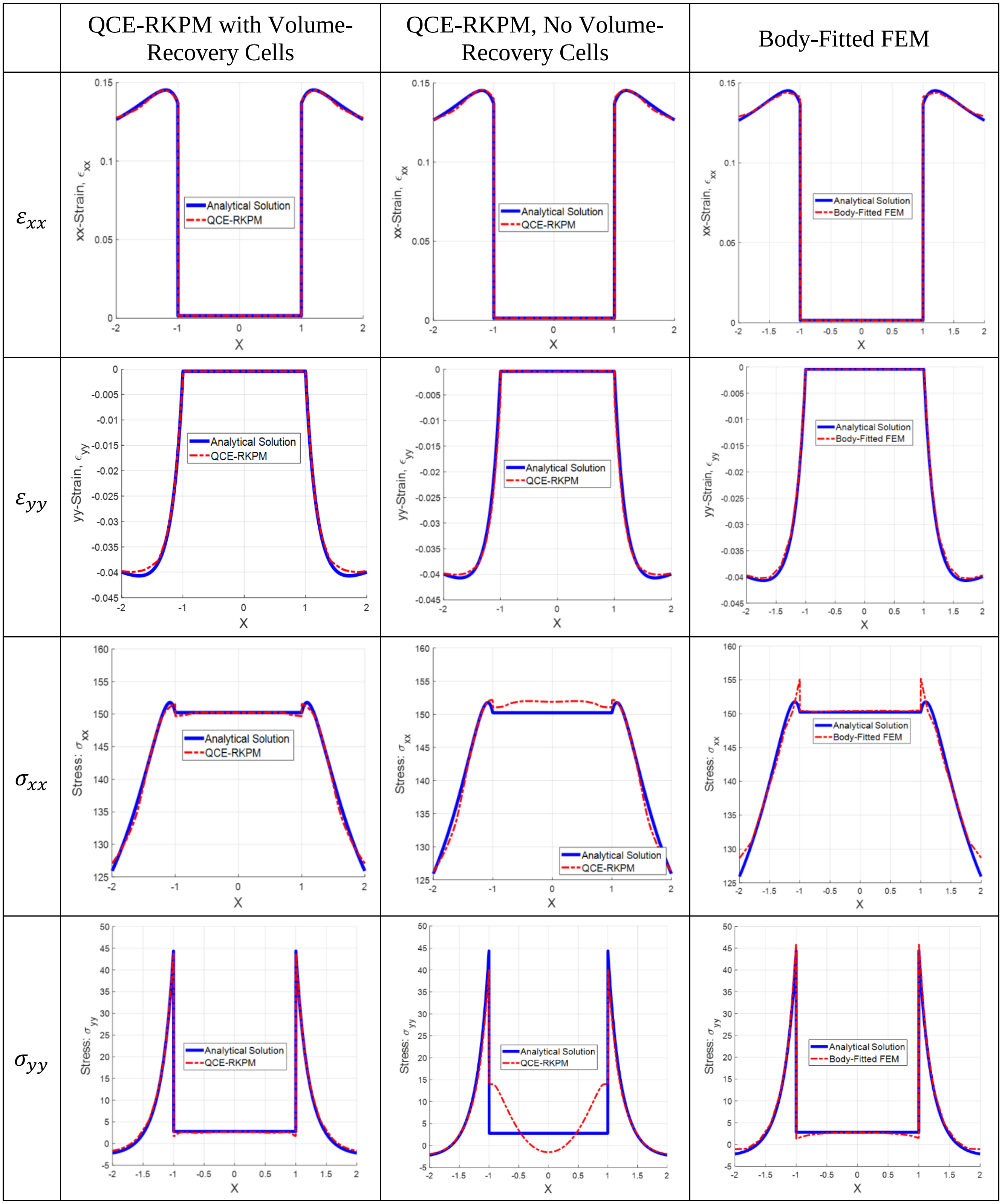

Figure 19: Cross-sections of the normal strains and stresses along $y = 0$ obtained by QCE-RKPM and Body-fitted FEM with discretizations in Figure 16

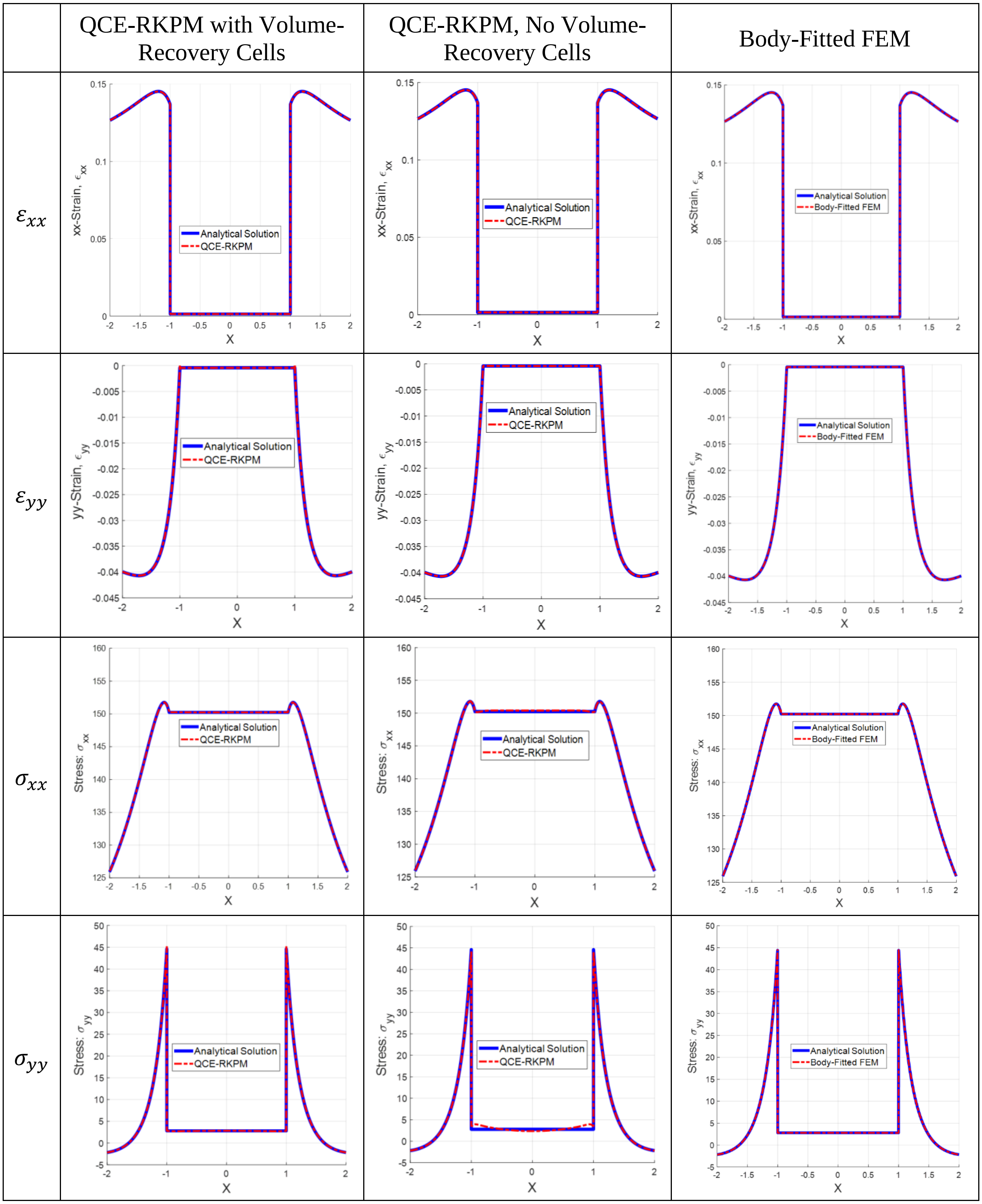

Figure 20: Cross-sections of the normal strains and stresses along $y = 0$ for the most refined discretization ($h^{+}_{\text{Intf}} \approx h^{-} = 0.0125$)

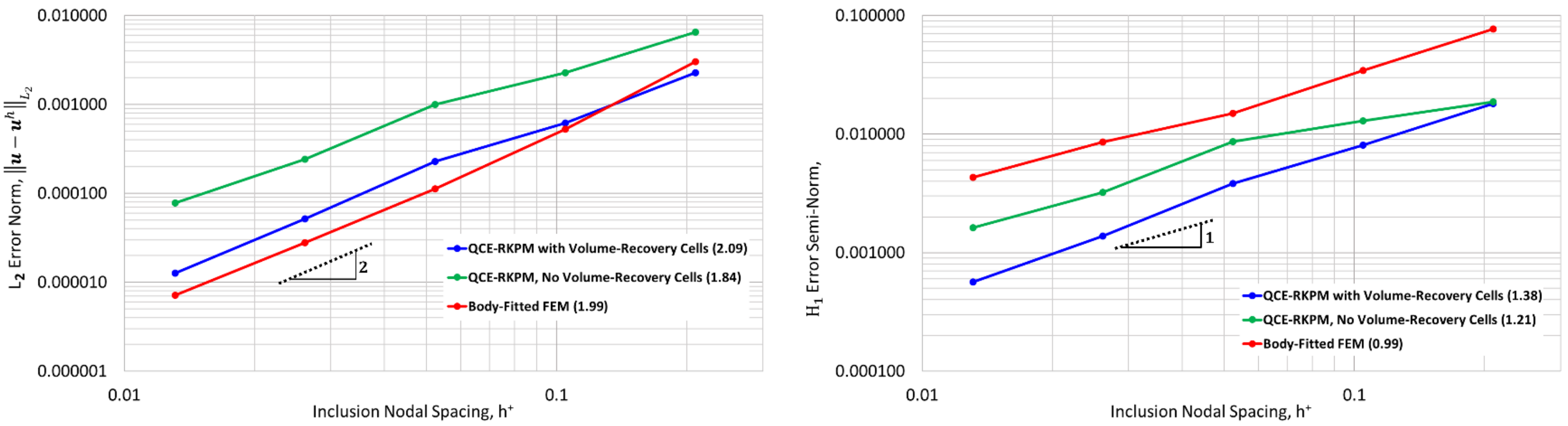

Figure 21: Convergence plots for the $L_2$ error norm and the $H_1$ error semi-norm

### 5.3. Multiple inclusion microstructure subjected to combined tension and shear loading

Our last example is of a heterogeneous microstructure with multiple, arbitrarily shaped interfaces, some of which contain concave features. Figure 22 shows a heterogeneous material consisting of randomly sized circular inclusions which is subjected to a combined tension and shear displacement of $\boldsymbol{g} = [0.02, 0.02]$. We assign the material properties to be $E^- = 1{,}000$, $\nu^- = 0.3$, $E^+ = 100{,}000$, $\nu^+ = 0.3$.

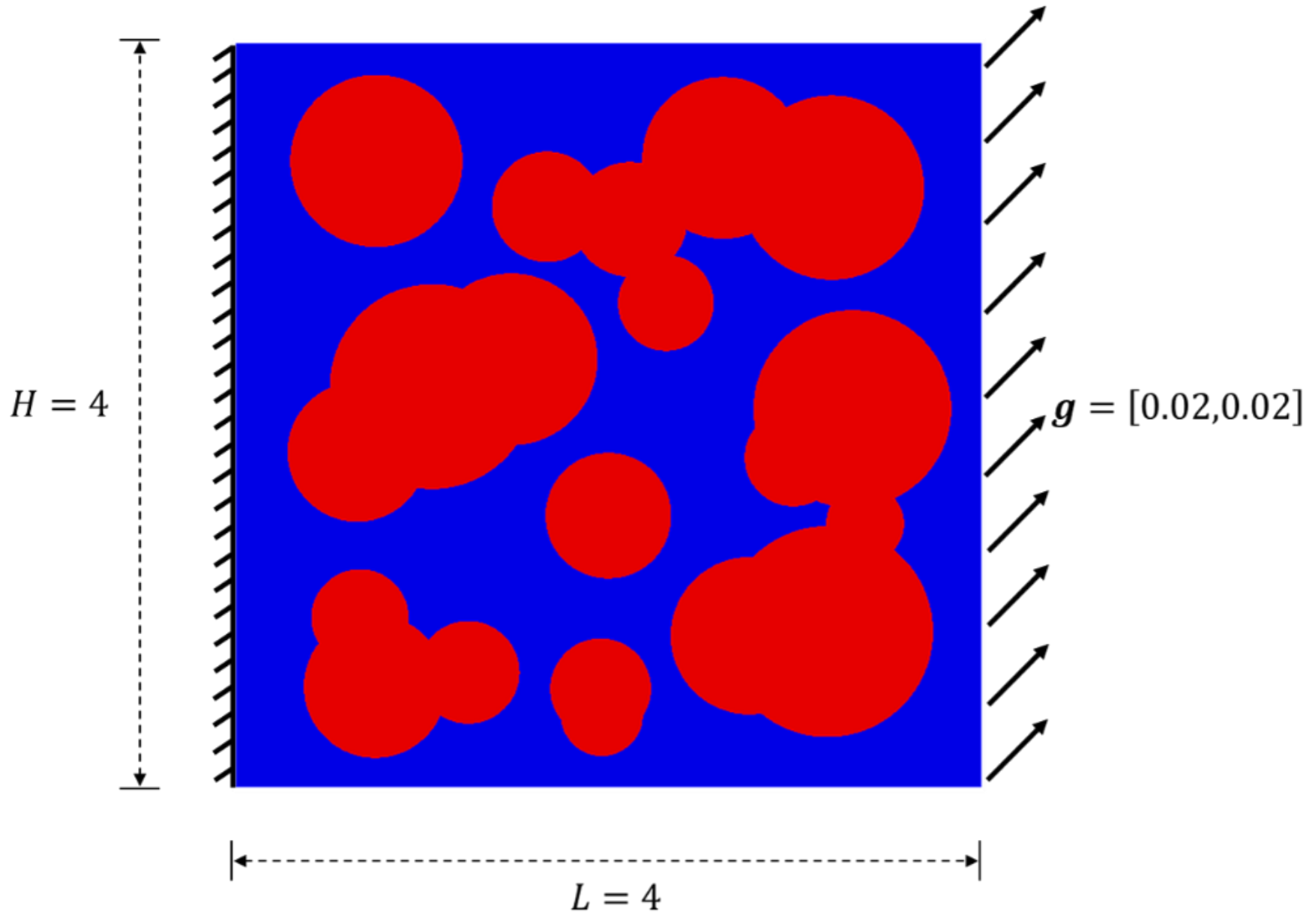

Figure 22: Combined tensile and shear loading of a heterogeneous microstructure

In this study, QCE-RKPM results obtained from a coarse smoothing-cell discretization and two finer smoothing-cell discretizations with and without quadtree subdivisions (Figure 23) are compared with the reference solution obtained from a body-fitted FEM model (Figure 25). As shown in Figure 24, all QCE-RKPM cases possess background integration smoothing cells that are non-conforming near the interface. The QCE-RKPM discretizations in Figure 23(a), (b), and (c) are with 709, 1,639, and 1,839 nodes, respectively, and the FEM discretization in Figure 25 is with 424,063 nodes. All QCE-RKPM discretizations utilize the volume-recovery integration smoothing cells for enhanced accuracy. The FEM reference solution was run in Abaqus CAE [76] and post-processed in MATLAB. The results displayed in Figure 26 show that QCE-RKPM can achieve similar results to a highly refined, body-fitted FEM solution without the effort of constructing a body-fitted discretization.

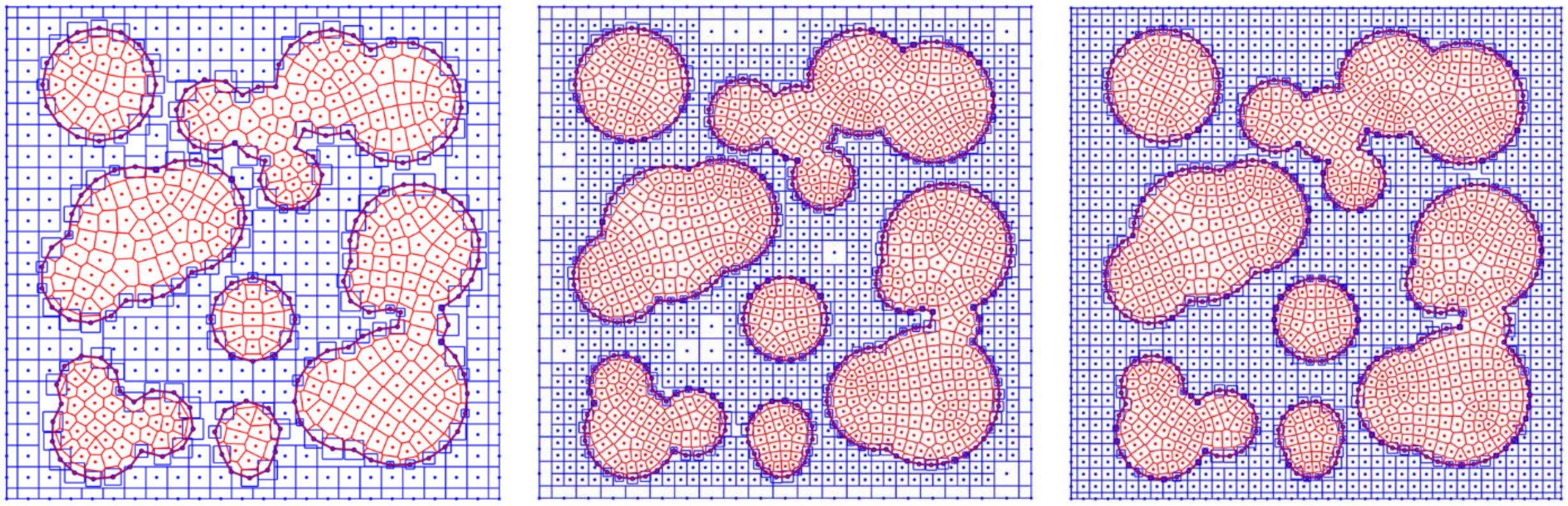

(a) (b) (c)

Figure 23: (a) $h^- \approx h^+ \approx 0.175$ with 709 nodes; (b) $h^- \approx 0.2, h^+ \approx 0.1$ with 1,639 nodes; (c) $h^- \approx h^+ \approx 0.1$ with 1,839 nodes

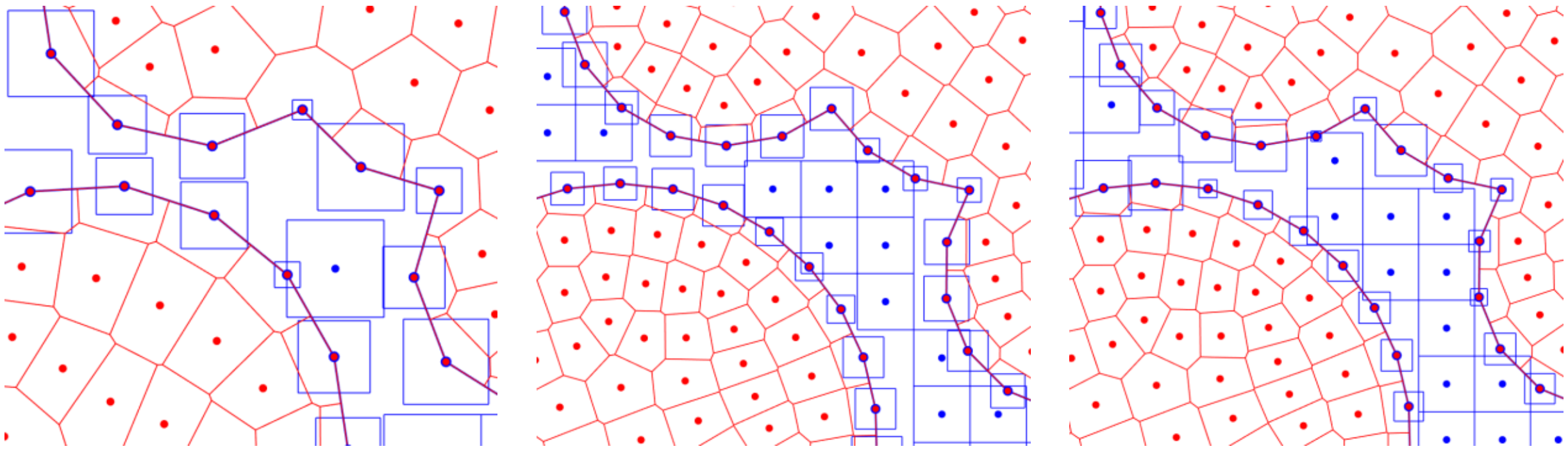

(a) (b) (c)

Figure 24: Magnified views of (a) Figure 23(a), (b) Figure 23(b), and (c) Figure 23(c)

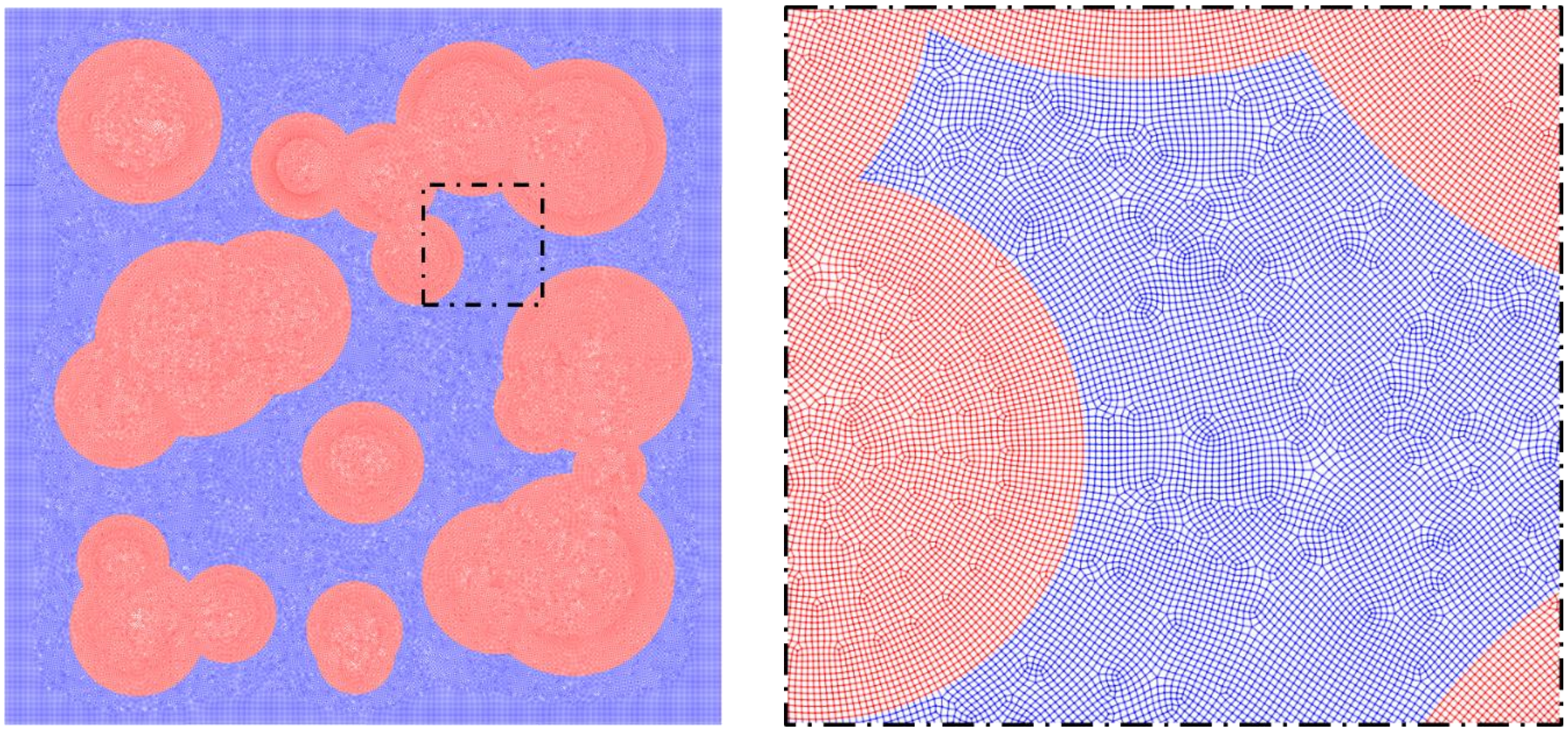

Figure 25: Conforming FEM discretization with $h^{-} = h^{+} \approx 0.00625$ with 424,063 nodes used as the reference solution

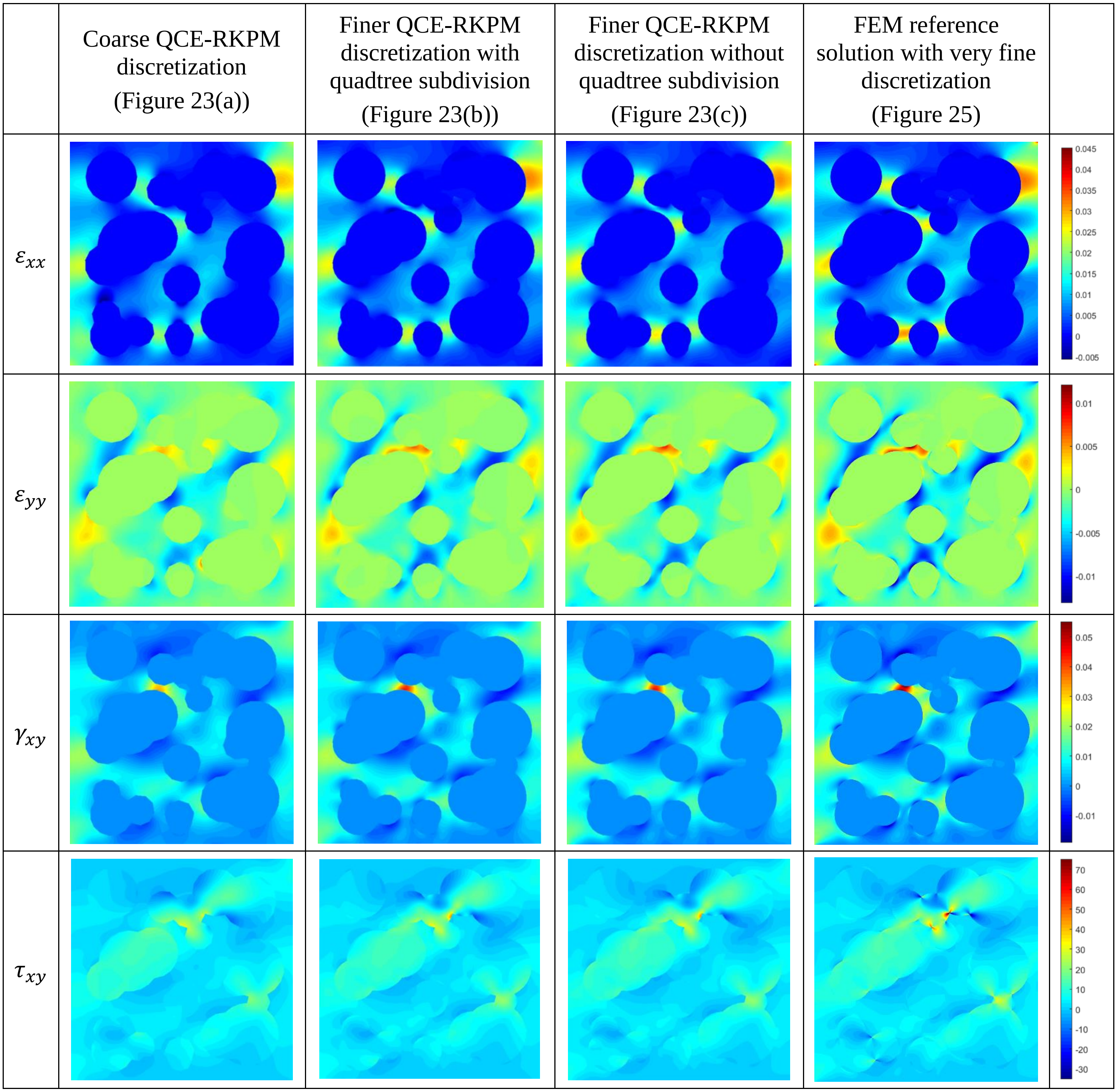

Figure 26: Strain and $\tau_{xy}$ fields for the QCE-RKPM discretizations shown in Figure 23 and the body-fitted FEM reference discretization shown in Figure 25

# 6. Summary

In this work, we proposed a quasi-conforming embedded reproducing kernel particle method (QCE-RKPM) for problems involving heterogeneous materials. The proposed formulation has demonstrated many advantages over the standard mesh-based immersed and embedded methods. The inclusion and matrix domains are discretized independently which avoids time-consuming conformal meshing. After immersing the inclusion discretization in a uniform matrix discretization, the fictitious domain, the source of stability and interpolation error problems in conventional immersed methods, is removed. Due to the non-conforming discretization, the VC correction is needed to meet first order exactness in the Galerkin approximation for optimal convergence. Additional interface smoothing cells with volume correction, while non-conforming, can be introduced in a straightforward way to further enhance solution accuracy. This ability to add nodes and smoothing cells where needed without concerns for mesh connectivity and conformity is a major advantage over mesh-based methods which require cell-cutting, cell-aggregation, or surrogate domain techniques. Finally, the inclusion nodes on the interface are shared with the matrix discretization to provide additional coercivity for a penalty-free Nitsche's method as well as ensuring the partition of unity in the approximation in each of the sub-domains. Thus, unlike most immersed and embedded methods, our formulation does not have any tunable parameters present to enforce interface compatibility.

QCE-RKPM was implemented with stabilized conforming/non-conforming nodal integration, variationally consistent correction, NSNI stabilization, and a line-of-sight algorithm. Several numerical examples were given which show that QCE-RKPM is a viable embedded method for modeling weak discontinuities in heterogeneous materials. The first numerical example shows that the proposed QCE-RKPM passes the piecewise linear patch test with weak discontinuities, and in problems with piecewise higher order solutions, it exhibits optimal convergence in the $L_2$ norm and superconvergence in the $H_1$ semi-norm. In a two-dimensional example of an infinite plate with a circular inclusion, QCE-RKPM with volume-recovery cells achieves accuracy comparable to body-fitted FEM and yields optimal convergence rates in the $L_2$ norm. It also shows that QCE-RKPM with or without volume-recovery cells outperforms body-fitted FEM in the $H_1$ semi-norm in both overall error and convergence rate. Finally, a two-dimensional example of a multi-inclusion microstructure was simulated using QCE-RKPM and was compared with a highly refined, body-fitted FEM mesh. The QCE-RKPM with much coarser discretizations provided similar results to the highly-refined, body-fitted FEM reference solution.

In summary, our method is distinct from standard, mesh-based immersed and embedded formulations. Unlike immersed methods in general, our formulation does not involve a fictitious domain and thus does not have a discretization ratio limitation. Unlike VCIMs, it presents a sharp discontinuity at the interface which captures the gradient jump. Unlike standard embedded methods, our formulation does not suffer from the small cut element problem, use surrogate domains, or aggregate elements.

**Acknowledgements**

The support of this work by the Sandia National Laboratories to UC San Diego under Contract Agreement 1655264 is greatly acknowledged. Sandia National Laboratories is a multi-mission laboratory managed and operated by National Technology and Engineering Solutions of Sandia, LLC, a wholly owned subsidiary of Honeywell International Inc., for the U.S. Department of Energy's National Nuclear Security Administration under contract DE-NA0003525. This paper describes objective technical results and analysis. Any subjective views or opinions that might be expressed in the paper do not necessarily represent the views of the U.S. Department of Energy or the United States Government. The support from National Science Foundation under Award Number CMMI-1826221 to University of California, San Diego is greatly appreciated.

**Appendix: Galerkin and Matrix Forms**

Let $\boldsymbol{u}^{h-}$ and $\boldsymbol{u}^{h+}$ be the RK approximations of $\boldsymbol{u}^{-}$ and $\boldsymbol{u}^{+}$ in $\Omega^{-}$ and $\Omega^{+}$, respectively. Substituting these approximations into the weak form of (29) and (30) yields the Galerkin form: Find $(\boldsymbol{u}^{h+}, \boldsymbol{u}^{h-}) \in H^1(\Omega^+) \times H^1(\Omega^-)$ such that $\forall(\delta\boldsymbol{u}^{h+}, \delta\boldsymbol{u}^{h-}) \in H^1(\Omega^+) \times H^1(\Omega^-)$

$$
\begin{aligned}
&\left(\boldsymbol{\varepsilon}(\delta\boldsymbol{u}^{h-}), \boldsymbol{\sigma}^{-}(\boldsymbol{u}^{h-})\right)_{\Omega^-} + \left(\delta\boldsymbol{u}^{h-}, \boldsymbol{n}^{+}\cdot\boldsymbol{\sigma}^{+}(\boldsymbol{u}^{h+})\right)_{\Gamma} - \left(\delta\boldsymbol{u}^{h-}, \boldsymbol{n}^{-}\cdot\boldsymbol{\sigma}^{-}(\boldsymbol{u}^{h-})\right)_{\partial\Omega_D} \\
&\quad - (\boldsymbol{\sigma}^{-}(\delta\boldsymbol{u}^{h-})\cdot\boldsymbol{n}^{-}, \boldsymbol{u}^{h-})_{\partial\Omega_D} + \beta(\delta\boldsymbol{u}^{h-}, \boldsymbol{u}^{h-})_{\partial\Omega_D} \\
&\quad = (\delta\boldsymbol{u}^{h-}, \boldsymbol{b}^{-})_{\Omega^-} + (\delta\boldsymbol{u}^{h-}, \boldsymbol{t})_{\partial\Omega_N} - (\boldsymbol{\sigma}^{-}(\delta\boldsymbol{u}^{h-})\cdot\boldsymbol{n}^{-}, \boldsymbol{g})_{\partial\Omega_D} \\
&\quad + \beta(\delta\boldsymbol{u}^{h-}, \boldsymbol{g})_{\partial\Omega_D},
\end{aligned}
\tag{76}
$$

$$
\left(\boldsymbol{\varepsilon}(\delta\boldsymbol{u}^{h+}), \boldsymbol{\sigma}^{+}(\boldsymbol{u}^{h+})\right)_{\Omega^+} - \left(\delta\boldsymbol{u}^{h+}, \boldsymbol{n}^{+}\cdot\boldsymbol{\sigma}^{+}(\boldsymbol{u}^{h+})\right)_{\Gamma} = (\delta\boldsymbol{u}^{h+}, \boldsymbol{b}^{+})_{\Omega^+}, \tag{77}
$$

Using VC-SNNI and SCNI both with NSNI stabilization for the background matrix and foreground inclusion domains, respectively, we arrive at our matrix form:

$$
\begin{aligned}
&\left[\int_{\Omega^-} \hat{\hat{\boldsymbol{B}}}_I^{-T} \boldsymbol{C}^{-}\, \hat{\boldsymbol{B}}_J^{-}\, d\Omega - \int_{\partial\Omega_D} \boldsymbol{\Psi}_I^{-T} \boldsymbol{\eta}^{-T} \boldsymbol{C}^{-}\, \tilde{\boldsymbol{B}}_J^{-}\, d\Gamma - \int_{\partial\Omega_D} \tilde{\boldsymbol{B}}_I^{-T} \boldsymbol{C}^{-}\, \boldsymbol{\eta}^{-}\, \boldsymbol{\Psi}_J^{-}\, d\Gamma \right. \\
&\left. + \beta \int_{\partial\Omega_D} \boldsymbol{\Psi}_I^{-T} \boldsymbol{\Psi}_J^{-}\, d\Gamma \right] \boldsymbol{d}_J^{-} + \int_{\Gamma} \boldsymbol{\Psi}_I^{-T} \boldsymbol{\eta}^{+T} \boldsymbol{C}^{+}\, \tilde{\boldsymbol{B}}_J^{+}\, d\Gamma\ \boldsymbol{d}_J^{+} \\
&= \int_{\Omega^-} \boldsymbol{\Psi}_I^{-T} \boldsymbol{b}^{-} d\Omega + \int_{\partial\Omega_N} \boldsymbol{\Psi}_I^{-T} \boldsymbol{t}\, d\Gamma - \int_{\partial\Omega_D} \tilde{\boldsymbol{B}}_I^{-T} \boldsymbol{C}^{-}\, \boldsymbol{\eta}^{-}\, \boldsymbol{g}\, d\Gamma + \beta \int_{\partial\Omega_D} \boldsymbol{\Psi}_I^{-T} \boldsymbol{g}\, d\Gamma,
\end{aligned}
\tag{78}
$$

$$\left[\int_{\Omega^+} \widehat{\boldsymbol{B}}_I^{+^T} \boldsymbol{C}^+ \, \widehat{\boldsymbol{B}}_J^+ \, d\Omega - \int_{\Gamma} \boldsymbol{\Psi}_I^{+^T} \boldsymbol{\eta}^{+^T} \boldsymbol{C}^+ \, \widetilde{\boldsymbol{B}}_J^+ \, d\Gamma\right] \boldsymbol{d}_J^+ = \int_{\Omega^+} \boldsymbol{\Psi}_I^{+^T} \boldsymbol{b}^+ d\Omega, \tag{79}$$

or:

$$\begin{bmatrix} \widehat{\boldsymbol{K}}_{IJ}^{--} - \breve{\boldsymbol{K}}_{IJ}^{--} - \breve{\boldsymbol{K}}_{IJ}^{--^T} + \boldsymbol{K}_{IJ}^{\beta--} & \boldsymbol{K}_{IJ}^{\Gamma-+} \\ \boldsymbol{0} & \widehat{\boldsymbol{K}}_{IJ}^{++} - \boldsymbol{K}_{IJ}^{\Gamma++} \end{bmatrix} \begin{bmatrix} \boldsymbol{d}_J^- \\ \boldsymbol{d}_J^+ \end{bmatrix} = \begin{bmatrix} \boldsymbol{f}_I^{b-} + \boldsymbol{f}_I^{t-} - \breve{\boldsymbol{f}}_I^{g-} + \boldsymbol{f}_I^{\beta-} \\ \boldsymbol{f}_I^{b+} \end{bmatrix}, \tag{80}$$

where:

$$\widehat{\boldsymbol{K}}_{IJ}^{--} = \int_{\Omega^-} \widehat{\widehat{\boldsymbol{B}}}_I^{-^T} \boldsymbol{C}^- \, \widehat{\boldsymbol{B}}_J^- \, d\Omega\,, \tag{81}$$

$$\breve{\boldsymbol{K}}_{IJ}^{--} = \int_{\partial\Omega_D} \boldsymbol{\Psi}_I^{-T} \boldsymbol{\eta}^{-T} \boldsymbol{C}^- \, \widetilde{\boldsymbol{B}}_J^- \, d\Gamma\,, \tag{82}$$

$$\boldsymbol{K}_{IJ}^{\beta--} = \beta \int_{\partial\Omega_D} \boldsymbol{\Psi}_I^{-T} \boldsymbol{\Psi}_J^- \, d\Gamma\,, \tag{83}$$

$$\boldsymbol{K}_{IJ}^{\Gamma-+} = \int_{\Gamma} \boldsymbol{\Psi}_I^{-T} \boldsymbol{\eta}^{+^T} \boldsymbol{C}^+ \, \widetilde{\boldsymbol{B}}_J^+ \, d\Gamma\,, \tag{84}$$

$$\boldsymbol{f}_I^{b-} = \int_{\Omega^-} \boldsymbol{\Psi}_I^{-T} \boldsymbol{b}^- d\Omega\,, \tag{85}$$

$$\boldsymbol{f}_I^{t-} = \int_{\partial\Omega_N} \boldsymbol{\Psi}_I^{-T} \boldsymbol{t} \, d\Gamma\,, \tag{86}$$

$$\breve{\boldsymbol{f}}_I^{g-} = \int_{\partial\Omega_D} \widetilde{\boldsymbol{B}}_I^{-^T} \boldsymbol{C}^- \, \boldsymbol{\eta}^- \, \boldsymbol{g} \, d\Gamma\,, \tag{87}$$

$$\boldsymbol{f}_I^{\beta-} = \beta \int_{\partial\Omega_D} \boldsymbol{\Psi}_I^{-T} \boldsymbol{g} \, d\Gamma\,, \tag{88}$$

$$\widehat{\boldsymbol{K}}_{IJ}^{++} = \int_{\Omega^+} \widehat{\boldsymbol{B}}_I^{+^T} \boldsymbol{C}^+ \, \widehat{\boldsymbol{B}}_J^+ \, d\Omega\,, \tag{89}$$

$$\boldsymbol{K}_{IJ}^{\Gamma++} = \int_{\Gamma} \boldsymbol{\Psi}_I^{+T} \boldsymbol{\eta}^{+T} \boldsymbol{C}^{+} \widetilde{\boldsymbol{B}}_J^{+} \, d\Gamma , \tag{90}$$

$$\boldsymbol{f}_I^{b+} = \int_{\Omega^+} \boldsymbol{\Psi}_I^{+T} \boldsymbol{b}^{+} d\Omega . \tag{91}$$

and where $\boldsymbol{d}_J^-$ and $\boldsymbol{d}_J^+$ are the background and foreground generalized nodal displacements, respectively. Again, $\hat{\hat{\boldsymbol{B}}}_I$ is a VC-corrected and NSNI stabilized smoothed strain-displacement matrix, $\hat{\boldsymbol{B}}_I$ is a NSNI stabilized smoothed strain-displacement matrix, $\widetilde{\boldsymbol{B}}_I$ is a smoothed strain-displacement matrix, $\boldsymbol{\Psi}_I = \Psi_I \mathbf{I}$ is the multi-dimensional shape function matrix, $\mathbf{I}$ is the identity matrix, and $\boldsymbol{\eta}$ is the matrix of outward normal vector components on either the interface $\Gamma$ or the outer boundary $\partial\Omega$.

**Remark A.1**

*Note that the smoothed strain-displacement matrices used for domain integration have been reused in the contour integrals for computational efficiency. From numerical experiments, we have determined that using the smoothed gradient in the contour integrals works just as well as using the direct gradient of the RK shape function.*

# 7. References

[1] C. S. Peskin, "Flow Patterns Around Heart Valves: A Numerical Method," *Journal of Computational Physics*, vol. 10, no. 2, pp. 252-271, 1972.

[2] C. S. Peskin, "The immersed boundary method," *Acta Numerica*, vol. 11, pp. 479-517, 2002.

[3] D. Boffi and L. Gastaldi, "A finite element approach for the immersed boundary method," *Computers & Structures,* vol. 81, no. 8-11, pp. 491-501, 2003.

[4] D. Boffi, L. Gastaldi and L. Heltai, "Numerical stability of the finite element immersed boundary method," *Mathematical Models and Methods in Applied Sciences,* vol. 17, no. 10, pp. 1479-1505, 2007.

[5] B. E. Griffith and X. Luo, "Hybrid finite difference/finite element immersed boundary method," *International Journal for Numerical Methods in Biomedical Engineering,* vol. 33, no. 12, p. e2888, 2017.

[6] F. Sotiropoulos and X. Yang, "Immersed boundary methods for simulating fluid–structure interaction," *Progress in Aerospace Sciences,* vol. 65, pp. 1-21, 2014.

[7] W. Kim and H. Choi, "Immersed boundary methods for fluid-structure interaction: A review," *International Journal of Heat and Fluid Flow,* vol. 75, pp. 301-309, 2019.

[8] Y. Bazilevs, K. Kamran, G. Moutsanidis, D. Benson and E. Oñate, "A new formulation for air-blast fluid–structure interaction using an immersed approach. Part I: basic methodology and FEM-based simulations," *Computational Mechanics,* vol. 60, pp. 83-100, 2017.

[9] Y. Bazilevs, G. Moutsanidis, J. Bueno, K. Kamran, D. Kamensky, M. C. Hillman, H. Gomez and J.-S. Chen, "A new formulation for air-blast fluid–structure interaction using an immersed approach: part II—coupling of IGA and meshfree discretizations," *Computational Mechanics,* vol. 60, no. 1, pp. 101-116, 2017.

[10] J. A. Cottrell, T. J. Hughes and Y. Bazilevs, Isogeometric Analysis: Toward Integration of CAD and FEA, New York: John Wiley & Sons, 2009.

[11] M. Behzadinasab, G. Moutsanidis, N. Trask, J. T. Foster and Y. Bazilevs, "Coupling of IGA and peridynamics for air-blast fluid-structure interaction using an immersed approach," *Forces in Mechanics,* vol. 4, p. 100045, 2021.

[12] R. Glowinsky, T.-W. Pan, T. I. Hesla, D. D. Joseph and J. Periaux, "A distributed Lagrange multiplier/fictitious domain method for the simulation of flow around moving rigid bodies: application to particulate flow," *Computer Methods in Applied Mechanics and Engineering,* vol. 184, no. 2-4, pp. 241-267, 2000.

[13] I. Babuška, "The finite element method with Lagrangian multipliers," *Numerische Mathematik,* vol. 20, no. 3, pp. 179-192, 1973.

[14] D. Boffi, F. Brezzi and M. Fortin, Mixed Finite Element Methods and Applications, vol. 44, Heidelberg: Springer, 2013.

[15] L. Zhang, A. Gerstenberger, X. Wang and W. K. Liu, "Immersed finite element method," *Computer Methods in Applied Mechanics and Engineering,* vol. 193, no. 21-22, pp. 2051-2067, 2004.

[16] L. T. Zhang and M. Gay, "Immersed finite element method for fluid-structure interactions," *Journal of Fluids and Structures,* vol. 23, no. 6, pp. 839-857, 2007.

[17] X. Wang and L. T. Zhang, "Modified immersed finite element method for fully-coupled fluid–structure interactions," *Computer Methods in Applied Mechanics and Engineering,* vol. 267, pp. 150-169, 2013.

[18] W. K. Liu, S. Jun and Y. F. Zhang, "Reproducing kernel particle methods," *International Journal of Numerical Methods in Fluids,* vol. 20, no. 8-9, pp. 1081-1106, 1995.

[19] J.-S. Chen, C. Pan, C. Wu and W. K. Liu, "Reproducing kernel particle methods for large deformation analysis of nonlinear structures," *Computer Methods in Applied Mechanics and Engineering,* vol. 139, no. 1-4, pp. 195-227, 1996.

[20] X. Wang and L. T. Zhang, "Interpolation functions in the immersed boundary and finite element methods," *Computational Mechanics,* vol. 45, no. 4, pp. 321-334, 2010.

[21] T.-H. Huang, J.-S. Chen, M. R. Tupek, F. N. Beckwith, J. J. Koester and H. E. Fang, "A variational multiscale immersed meshfree method for heterogeneous materials," *Computational Mechanics,* vol. 67, no. 4, pp. 1059-1097, 2021.

[22] T. J. Hughes, L. P. Franca and M. Balestra, "A new finite element formulation for computational fluid dynamics: V. Circumventing the Babuška-Brezzi condition: A stable Petrov-Galerkin formulation of the Stokes problem accommodating equal-order interpolations," *Computer Methods in Applied Mechanics and Engineering,* vol. 59, no. 1, pp. 85-99, 1986.

[23] T. J. Hughes, G. R. Feijóo, L. Mazzei and J.-B. Quincy, "The variational multiscale method—a paradigm for computational mechanics," *Computational Methods in Applied Mechanics and Engineering,* vol. 166, no. 1-2, pp. 3-24, 1998.

[24] T.-H. Huang, J.-S. Chen, M. R. Tupek, F. N. Beckwith and H. E. Fang, "A variational multiscale immersed meshfree method for fluid structure interactive systems involving shock waves," *Computer Methods in Applied Mechanics and Engineering,* vol. 389, p. 114396, 2022.

[25] B. E. Griffith and N. A. Patankar, "Immersed Methods for Fluid–Structure Interaction," *Annual Review of Fluid Mechanics,* vol. 52, pp. 421-448, 2020.

[26] J. Wang, G. Zhou, M. Hillman, A. Madra, Y. Bazilevs, J. Du and K. Su, "Consistent immersed volumetric Nitsche methods for composite analysis," *Computer Methods in Applied Mechanics and Engineering,* vol. 385, p. 114042, 2021.

[27] A. Hansbo and P. Hansbo, "An unfitted finite element method, based on Nitsche's method, for elliptic interface problems," *Computer Methods in Applied Mechanics and Engineering,* vol. 191, no. 47-48, pp. 5537-5552, 2002.

[28] E. Burman, S. Claus, P. Hansbo, M. G. Larson and A. Massing, "CutFEM: Discretizing geometry and partial differential equations," *International Journal for Numerical Methods in Engineering,* vol. 104, no. 7, pp. 472-501, 2015.

[29] S. Soghrati, A. M. Aragón, C. A. Duarte and P. H. Geubelle, "An interface-enriched generalized FEM for problems with discontinuous gradient fields," *International Journal for Numerical Methods in Engineering,* vol. 89, no. 8, pp. 991-1008, 2012.

[30] S. Soghrati, "Hierarchical interface-enriched finite element method: An automated technique for mesh-independent simulations," *Journal of Computational Physics,* vol. 275, pp. 41-52, 2014.

[31] A. M. Aragón and A. Simone, "The Discontinuity-Enriched Finite Element Method," *International Journal for Numerical Methods in Engineering,* vol. 112, no. 11, pp. 1589-1613, 2017.

[32] E. Burman and P. Hansbo, "Fictitious domain finite element methods using cut elements: II. A stabilized Nitsche method," *Applied Numerical Mathematics,* vol. 62, no. 4, pp. 328-341, 2012.

[33] P. Hansbo, M. G. Larson and S. Zahedi, "A cut finite element method for a Stokes interface problem," *Applied Numerical Mathematics,* vol. 85, pp. 90-114, 2014.

[34] E. Burman and P. Hansbo, "Fictitious domain methods using cut elements: III. A stabilized Nitsche method for Stokes' problem," *ESAIM: Mathematical Modelling and Numerical Analysis,* vol. 48, no. 3, pp. 859-874, 2014.

[35] H. M. Mourad, J. Dolbow and I. Harari, "A bubble-stabilized finite element method for Dirichlet constraints on embedded interfaces," *International Journal for Numerical Methods in Engineering,* vol. 69, no. 4, pp. 772-793, 2007.

[36] J. E. Dolbow and L. P. Franca, "Residual-free bubbles for embedded Dirichlet problems," *Computer Methods in Applied Mechanics and Engineering,* vol. 197, no. 45-48, pp. 3751-3759, 2008.

[37] C. Annavarapu, M. Hautefeuille and J. E. Dolbow, "A robust Nitsche’s formulation for interface problems," *Computer Methods in Applied Mechanics and Engineering,* Vols. 225-228, pp. 44-54, 2012.

[38] W. Jiang, C. Annavarapu, J. E. Dolbow and I. Harari, "A robust Nitsche's formulation for interface problems with spline-based finite elements," *International Journal for Numerical Methods in Engineering,* vol. 104, no. 7, pp. 676-696, 2015.

[39] J. Parvizian, A. Düster and E. Rank, "Finite cell method: h- and p-extension for embedded domain problems in solid mechanics," *Computational Mechanics,* vol. 41, no. 1, pp. 121-133, 2007.

[40] A. Düster, J. Parvizian, Z. Yang and E. Rank, "The finite cell method for three-dimensional problems of solid mechanics," *Computer Methods in Applied Mechanics and Engineering,* vol. 197, pp. 3768-3782, 2008.

[41] D. Schillinger and M. Ruess, "The Finite Cell Method: A Review in the Context of Higher-Order Structural Analysis of CAD and Image-Based Geometric Models," *Archives of Computational Methods in Engineering,* vol. 22, no. 3, pp. 391-455, 2015.

[42] D. Kamensky, M.-C. Hsu, D. Schillinger, J. A. Evans, A. Aggarwal, Y. Bazilevs, M. S. Sacks and T. J. Hughes, "An immersogeometric variational framework for fluid–structure interaction: Application to bioprosthetic heart valves," *Computer Methods in Applied Mechanics and Engineering,* vol. 284, pp. 1005-1053, 2015.

[43] A. Main and G. Scovazzi, "The shifted boundary method for embedded domain computations. Part I: Poisson and Stokes problems," *Journal of Computational Physics,* vol. 372, pp. 972-995, 2018.

[44] A. Main and G. Scovazzi, "The shifted boundary method for embedded domain computations. Part II: Linear advection–diffusion and incompressible Navier–Stokes equations," *Journal of Computational Physics,* vol. 372, pp. 996-1026, 2018.

[45] K. Li, N. M. Atallah, G. A. Main and G. Scovazzi, "The Shifted Interface Method: A flexible approach to embedded interface computations," *International Journal for Numerical Methods in Engineering,* vol. 121, no. 3, pp. 492-518, 2020.

[46] J. Cheung, M. Perego, P. Bochev and M. Gunzburger, "Optimally accurate higher-order finite element methods for polytopial approximations of domains with smooth boundaries," *Mathematics of Computation,* vol. 88, no. 319, pp. 2187-2219, 2019.

[47] J. Cheung, M. Gunzburger, P. Bochev and M. Perego, "An optimally convergent higher-order finite element coupling method for interface and domain decomposition problems,"

*Results in Applied Mathematics,* vol. 6, p. 100094, 2020.

[48] N. M. Atallah, C. Canuto and G. Scovazzi, "The high-order Shifted Boundary Method and its analysis," *Computer Methods in Applied Mechanics and Engineering,* vol. 394, p. 114885, 2022.

[49] L. Nouveau, M. Ricchiuto and G. Scovazzi, "High-order gradients with the shifted boundary method: an embedded enriched mixed formulation for elliptic PDEs," *Journal of Computational Physics,* vol. 398, p. 108898, 2019.

[50] S. Badia, F. Verdugo and A. F. Martín, "The aggregated unfitted finite element method for elliptic problems," *Computer Methods in Applied Mechanics and Engineering,* vol. 336, pp. 533-553, 2018.

[51] S. Badia, A. F. Martin and F. Verdugo, "Mixed aggregated finite element methods for the unfitted discretization of the Stokes problem," *SIAM Journal on Scientific Computing,* vol. 40, no. 6, pp. B1541-B1576, 2018.

[52] S. Badia, E. Neiva and F. Verdugo, "Linking ghost penalty and aggregated unfitted methods," *Computer Methods in Applied Mechanics and Engineering,* vol. 388, p. 114232, 2022.

[53] A. Johansson and M. G. Larson, "A high order discontinuous Galerkin Nitsche method for elliptic problems with fictitious boundary," *Numerische Mathematik,* vol. 123, no. 4, pp. 607-628, 2013.

[54] W. Qiu, M. Solano and P. Vega, "A High Order HDG Method for Curved-Interface Problems Via Approximations from Straight Triangulations," *Journal of Scientific Computing,* vol. 69, no. 3, pp. 1384-1407, 2016.

[55] P. Huang, H. Wu and Y. Xiao, "An unfitted interface penalty finite element method for elliptic interface problems," *Computer Methods in Applied Mechanics and Engineering,* vol. 323, pp. 439-460, 2017.

[56] E. Burman, P. Hansbo and M. G. Larson, "CutFEM based on extended finite element spaces," *Numerische Mathematik,* vol. 152, no. 2, pp. 331-369, 2022.

[57] J.-S. Chen, M. Hillman and M. Ruter, "An arbitrary order variationally consistent integration for Galerkin meshfree methods," *International Journal for Numerical Methods in Engineering,* vol. 95, no. 5, pp. 387-418, 2013.

[58] D. Wang, Y. Sun and L. Li, "A Discontinuous Galerkin Meshfree Modeling of Material Interface," *Computer Modeling in Engineering & Sciences,* vol. 45, no. 1, pp. 57-82, 2009.

[59] T. Belytschko, J.-S. Chen and M. Hillman, Meshfree and Particle Methods, Hoboken, NJ: Wiley Publishing Company, forthcoming.

[60] T. Belytschko, Y. Y. Lu and L. Gu, "Element-free Galerkin methods," *International Journal for Numerical Methods in Engineering,* vol. 37, no. 2, pp. 229-256, 1994.

[61] P. Krysl and T. Belytschko, "Element-free Galerkin method: Convergence of the continuous and discontinuous shape functions," *Computer Methods in Applied Mechanics and Engineering,* vol. 148, no. 3-4, pp. 257-277, 1997.

[62] T. Belytschko, Y. Krongauz, D. Organ, M. Fleming and P. Krysl, "Meshless methods: an overview and recent developments," *Computer Methods in Applied Mechanics and Engineering,* vol. 139, no. 1-4, pp. 3-47, 1996.

[63] D. Organ, M. Fleming, T. Terry and T. Belytschko, "Continuous meshless approximations for nonconvex bodies by diffraction and transparency," *Computational Mechanics,* vol. 18, pp. 225-235, 1996.

[64] J. Dolbow and T. Belytschko, "Numerical integration of the Galerkin weak form in meshfree methods," *Computational Mechanics,* vol. 23, no. 3, pp. 219-230, 1999.

[65] J.-S. Chen, C.-T. Wu, S. Yoon and Y. You, "A stabilized conforming nodal integration for Galerkin meshfree methods," *International Journal for Numerical Methods in Engineering,* vol. 50, no. 2, pp. 435-466, 2001.

[66] M. Hillman and J.-S. Chen, "An accelerated, convergent, and stable nodal integration in Galerkin meshfree methods for linear and nonlinear mechanics," *International Journal for Numerical Methods in Engineering,* vol. 107, no. 7, pp. 603-630, 2016.

[67] J.-S. Chen, M. Hillman and S.-W. Chi, "Meshfree Methods: Progress Made After 20 Years," *Journal of Engineering Mechanics,* vol. 143, no. 4, p. 04017001, 2017.

[68] J.-S. Chen and Y. Wu, "Stability in Lagrangian and semi-Lagrangian reproducing kernel discretizations using nodal integration in nonlinear solid mechanics," in *Computational Methods in Applied Sciences,* Dordrecht, 2007.

[69] P.-C. Guan, J.-S. Chen, Y. Wu, H. Teng, J. Gaidos, K. Hofstetter and M. Alsaleh, "Semi-Lagrangian reproducing kernel formulation and application to modeling earth moving operations," *Mechanics of Materials,* vol. 41, no. 6, pp. 670-683, 2009.

[70] J.-S. Chen, W. Hu, M. A. Puso, Y. Wu and X. Zhang, "Strain smoothing for stabilization and regularization of Galerkin meshfree method," in *Meshfree Methods for Partial Differential Equations III,* vol. 57, Berlin, Springer, 2007, pp. 57-76.

[71] M. A. Puso, J.-S. Chen, E. Zywicz and W. Elmer, "Meshfree and finite element nodal integration methods," *International Journal for Numerical Methods in Engineering,* vol. 74, no. 3, pp. 416-446, 2008.

[72] S. Li and W. K. Liu, "Reproducing kernel hierarchical partition of unity, part I—formulation and theory," *International Journal for Numerical Methods in Engineering,* vol. 45, no. 3, pp. 251-288, 1999.

[73] S. Li and W. K. Liu, "Reproducing kernel hierarchical partition of unity, part II—applications," *International Journal for Numerical Methods in Engineering,* vol. 45, no. 3, pp. 289-317, 1999.

[74] J.-S. Chen, X. Zhang and T. Belytschko, "An implicit gradient model by a reproducing kernel strain regularization in strain localization problems," *Computer Methods in Applied Mechanics and Engineering,* vol. 193, no. 27-29, pp. 2827-2844, 2004.

[75] M. L. Kachanov, B. Shafiro and I. Tsukrov, Handbook of Elasticity Solutions, Dordrecht: Springer Science & Business Media, 2003.

[76] Simulia Corp., "Abaqus/CAE User's Guide," Simulia Corp., 2016. [Online]. Available: http://130.149.89.49:2080/v2016/books/usi/default.htm. [Accessed 10 June 2020].